------------------- prb.tex follows --------------------
\documentstyle [12pt]{article}
\newcommand{\cc}{CeCu$_2$Si$_2$ }
\newcommand{\ga}{\Gamma\alpha}
\newcommand{\gp}{\Gamma^{\prime}}
\newcommand{\ap}{\alpha^{\prime}}
\newcommand{\kq}{\vec k +\vec q}
\newcommand{\kp}{\vec k^{\prime}}
\begin{document}
\title{A Superconducting Instability in the Infinite-U Anderson Lattice
in the Presence of Crystal Electric Fields}
\author{B. R. Trees$^*$\\
Department of Physics\\
The Ohio State University\\
174 W. 18th Ave.\\
Columbus, OH 43210}
\date{\today}
\maketitle

\setlength{\baselineskip}{24pt}

\begin{center}
{\bf Abstract}
\end{center}
\vspace{0.2cm}
We report evidence of a superconducting instability (of $T_{1g}$
symmetry) in the infinite-U Anderson lattice in the presence of crystal fields
of
cubic symmetry.  We assume a lattice of $4f$ sites, each with a total angular
momentum of
$J=5/2$ that is split by crystal fields into a low-lying doublet of $\Gamma_7$
symmetry and an excited quartet of $\Gamma_8$ symmetry.  Slave Bosons on the
$4f$ sites create and destroy $4f^0$ configurations and Lagrange multipliers at
each
$4f$ site enforce the occupancy constraint due to the infinite Coulomb
repulsion.
Quasiparticle interactions
are due to exchange of $4f$ density fluctuations, which are represented by
fluctuations in the
slave Bosons and Lagrange multipliers.
We use the so-called analytic tetrahedron method
to calculate the dressed (to order 1/N) Boson Green functions.  In weak
couping, the exchange of the
dressed Bosons gives rise to a superconducting instability of $T_{1g}$,
$xy(x^2-y^2)$,
symmetry.  The $A_{1g}$, ``s-wave'', channel has strongly repulsive
interactions and hence
no pairing instability.  The $T_{2g}$ channel exhibits weakly repulsive
interactions.
Average quasiparticle interactions in the $E_g$, $x^2-y^2$, $3z^2-r^2$, channel
fluctuate
strongly as a function of the number of tetrahedra used to calculate the
Bosonic Green functions,
lending only weak evidence for an instability of $E_g$ symmetry.

\smallskip
\smallskip


\pagebreak

\begin{center}
{\Large  I. INTRODUCTION}
\end{center}
\vspace{0.4cm}
This paper is concerned with the effects of crystal electric
fields on quasiparticle interactions in the Ce based heavy Fermion
superconductor CeCu$_2$Si$_2$ ($T_c\approx$0.6 K\cite{Steglich 79}).
In general, the heavy Fermion materials are examples of systems
exhibiting strong correlations among the constituent particles
\cite{Grewe 90} \cite{Stewart 84}, \cite{Lee 86},
\cite{Weslau 92}, \cite{Fulde 88}, \cite{Rainer 88}.
Generally, the compounds are comprised of intermetallics and rare-earth
or actinide atoms (such as uranium or cerium) with a strong on-site
Coulomb repulsion.  This large electrostatic energy arises from the
extremely localized nature of the 4f or 5f wavefunctions in the solid
and markedly influences the electron occupation at these ``rare-earth''
sites.  When hybridization between a rare-earth electron and a
conduction electron is allowed, the physics of this strong interaction
is communicated to the solid at large, giving rise to a metal of
strongly correlated, interacting electrons.  In such a system one might
expect to find a ground state manifesting collective properties of the
coupled rare-earth and conduction electrons, e. g. superconductivity or
magnetism.  Indeed such ground states are seen.  There are also heavy
Fermion systems that apparently retain a metallic state down to zero
temperature.
To-date there are six known heavy Fermion
superconductors, all containing either cerium or uranium.


Our work is based on the infinite-U Anderson lattice, the details of which we
shall
discuss later.  For the experts, we mention here that we use
slave Boson operators
to create or destroy $4f^0$ configurations on the Ce sites, thereby avoiding
the cumbersome Hubbard operators in the hybridization piece of the Hamiltonian
\cite{Barnes 76},\cite{Read 83a},\cite{Coleman 84}.  There is also a Lagrange
multiplier
to enforce unit occupancy of the $4f$ multiplets at each Ce site.
Our work is novel in that we also include, at the Ce sites, crystal electric
fields of
cubic symmetry, which has the effect of splitting the spin-orbit
coupled ($J=5/2$) multiplet into a doublet (of $\Gamma_7$ symmetry) and a
quartet (of $\Gamma_8$ symmetry).
We take the $\Gamma_7$ doublet to be the ground multiplet, with a crystal field
splitting, $\Delta_{CEF}$, to the $\Gamma_8$ quartet that is much larger than
the Kondo temperature of the low-lying doublet, $T_{o7}$.  ($\Delta_{CEF}\gg
T_{o7}$)

Previous theoretical work has focused on understanding the heavy Fermion
compounds mainly
through the $SU(N)$
version of the periodic Anderson model\cite{Millis 86},\cite{Auerbach
86},\cite{Tesanovic 86},
\cite{Coleman 87},\cite{Lavagna 87}.
In the $SU(N)$ model, the $4f$ multiplet is $N$-fold degenerate, and the
(plane-wave) conduction
bands are assumed $N$-fold degenerate as well.  The matrix element, $V(\vec
k)$,
for hybridization between a conduction electron and a $4f$ electron, is taken
to be
isotropic in momentum space.

Within the $SU(N)$ model, Lavagna, Millis, and Lee\cite{Lavagna 87},
Auerbach and Levin\cite{Auerbach 86}, and Houghton, Read, and Won\cite{Houghton
87}
have studied
quasiparticle interactions due to the exchange of $4f$ density fluctuations.
The lowest
order diagrams contributing to the interactions are of order $1/N$, where $N$
is the
$4f$ multiplet degeneracy.  In Ce, in the absence of crystal field splitting,
the low-lying
$J=5/2$ multiplet is six-fold degenerate ($N=6$).  So it seems reasonable to
truncate the diagrams at order
$1/N$.  Lavagna, Millis, and Lee found such a spinless density exchange yielded
a $d$-wave
superconducting instability in the spin-singlet pairing channel.

F. C. Zhang and T. K. Lee\cite{Zhang 87} have performed a more realistic
calculation
(as far as heavy Fermion compounds are concerned) by including spin-orbit
coupling
at the Ce sites and by returning to the two-fold degenerate conduction states.
They included an anisotropic hybridization matrix element, of the
Coqblin-Schrieffer form\cite{Coqblin 69},
between conduction and $4f$ electrons .
These spin-orbit coupled ions are assumed to sit in an overall spherically
symmetric ``host'' (as
in a jellium model).
Unlike in the $SU(N)$ model,
in the even-parity pairing channel, Zhang and Lee found {\em no}
superconducting instabilities of $s$-wave, $d$-wave, or $g$-wave symmetry.  It
is worth noting
here that the mean field quasiparticle energy bands Zhang and Lee found are the
same as those
calculated by Zou and Anderson\cite{Zou 86}, who used the KKR scheme and
included spin-orbit
coupling on the $4f$ sites.

We have looked at density-fluctuation induced quasiparticle interactions in the
presence of crystal electric field splitting of the Ce $J=5/2$ multiplets in
the lattice.
We find the anisotropy due to cubic symmetry qualitatively and quantitatively
alters the
interactions in comparison to the results of Zhang and Lee.  In fact, we find
evidence for
a superconducting instability of $T_{1g}$ ($xy(x^2-y^2)$) symmetry in the
even-parity pairing channel.
We also find weaker evidence for an $E_g$ ($x^2-y^2$) pairing instability.  We
find {\em no}
instability in the $A_{1g}$ (the ``s-wave'' of cubic symmetry) channel, which
is not
surprising, given the strong Hubbard U repulsion built into the Anderson model.

The details of our calculation are presented here as follows. In Sec. II, we
briefly discuss
experimental evidence for the existence of crystal electric fields of cubic
symmetry in
\cc.  In Sec. III, we introduce the Hamiltonian for the infinite-U Anderson
lattice
in the presence of crystal fields.  The Hamiltonian formalism shall be retained
throughout this
paper, as opposed to functional integral techniques
\cite{Read 83a},\cite{Auerbach 86},\cite{Coleman 87},\cite{Lavagna
87},\cite{Read 83}.
In Sec. IV, we report the mean field properties of our Hamiltonian, including
the quasiparticle
energies and states.
In Sec. VI, we discuss our calculation of the dressed
slave Boson Green functions, which include the effects of particle-hole
excitations in the
hybridization coupled conduction and $4f$ electron system.  We also explain
(Secs. VII and VIII)
our use of the
so-called analytic tetrahedron method for performing the complicated three
dimensional
Brillouin zone integrals that arise in the evaluation of the particle-hole
diagrams.  Finally, in
Secs. X and XI, we present our results for the superconducting instabilities in
the presence of
cubic symmetry, which are based on the
Fermi surface average of the quasiparticle-quasiparticle scattering amplitude.

\begin{center}
{\Large II. CRYSTAL ELECTRIC FIELDS}
\end{center}
\vspace{0.4cm}

The first experimental evidence of crystal electric fields in \cc came from the
inelastic
neutron scattering data of
Horn {\em et al.}\cite{Horn 81}. The data show a clear peak at an energy of
31.5 meV
($\approx$ 360K) and a weaker peak at approximately 12 meV.  Since there was
also some
structure at 12 meV in the reference material LaCu$_2$Si$_2$ (which has no $4f$
electrons), it seems likely that the lower energy peak is due to phonons.  The
data could be fit, however, by assuming a crystal field structure in which the
6-fold
degenerate $J=5/2$ multiplets are split into three doublets, with excitation
energies
given by the measured energies of the inelastic peaks.  Such a three doublet
structure is
exactly what one would expect for a tetragonal crystal, like \cc.

Later, measurements of the specific heat at high temperatures \cite{Bredl 85},
however, did not
agree with this interpretation of the multiplet structure.  Bredl {\em et al.}
plotted
the $4f$ contribution to the specific heat of CeCu$_{2.2}$Si$_2$ by subtracting
off the
corresponding data for LaCu$_{2.2}$Si$_2$. They found they could not fit the
results
with two Schottky peaks, as would be expected for three crystal field doublets.
 Instead,
the excited magnetic states behaved as if there were a four-fold degenerate
multiplet
about 360 K above a ground state doublet.  This idea is also nicely
corroborated by Bredl
and co-workers' plot of the entropy as a function of temperature.  The rather
quick rise
of the entropy from a value of $R\ln$2 to $R\ln$6 at about 300 K would suggest
a quartet
structure rather than two doublets separated in energy by about 100K.
Such a structure would be appropriate in the presence of {\em cubic} symmetry.

Further evidence for {\em effective} cubic symmetry in \cc comes from the
dc susceptibility measurements by Steglich and co-workers on single crystal
\cc samples\cite{Steglich 85}.
They find a very weak anisotropy in the temperature
dependence of $\chi_{dc}$ for a magnetic field applied parallel or
perpendicular to
the c-axis of the unit cell.  Furthermore, the isotropy of the slope of the
upper critical magnetic field at $T_c$,
$H_{c2}^{\prime}(T_c)$\cite{Rauchschwalbe 87},
suggests that the
effective mass of the quasiparticles is isotropic.  Since the effective mass is
technically
a second-rank tensor, it can be isotropic only in the presence of cubic
symmetry.
In light of all this information, it is reasonable to assume that the small
peak in the neutron data at 12 meV is not due to crystal field excitations and
that there is effectively cubic symmetry at the cerium sites.

In the presence of cubic symmetry, the J=5/2 multiplet is split into a
doublet of $\Gamma$$_7$ symmetry and
a $\Gamma$$_8$ quartet \cite{Lea 62}.  The crystal field-split states
$|\ga\rangle$,
where $\alpha$ labels the degenerate states for a given multiplet, are a linear
combination
of the eigenstates of the z-component of the total angular momentum,
$|m\rangle$, where
$-5/2\leq m\leq 5/2$:
\begin{equation} \label{eq:71}
|\Gamma_7,+1\rangle=-\sqrt{\frac{1}{6}}|-5/2\rangle +
\sqrt{\frac{5}{6}}|3/2\rangle
\end{equation}
\begin{equation} \label{eq:7-1}
|\Gamma_7,-1\rangle=-\sqrt{\frac{1}{6}}|5/2\rangle +
\sqrt{\frac{5}{6}}|-3/2\rangle
\end{equation}
\begin{equation} \label{eq:82}
|\Gamma_8,+2\rangle=\sqrt{\frac{5}{6}}|5/2\rangle +
\sqrt{\frac{1}{6}}|-3/2\rangle
\end{equation}
\begin{equation} \label{eq:81}
|\Gamma_8,1\rangle=|1/2\rangle
\end{equation}
\begin{equation} \label{eq:8-1}
|\Gamma_8,-1\rangle=|-1/2\rangle
\end{equation}
\begin{equation} \label{eq:8-2}
|\Gamma_8,-2\rangle=\sqrt{\frac{5}{6}}|-5/2\rangle +
\sqrt{\frac{1}6}|3/2\rangle.
\end{equation}

{}From the neutron scattering and specific heat data we know that the ground
multiplet is the doublet.
Thus, without
addressing the true microscopic source of the crystal fields, we deduce a
multiplet
structure as shown in figure 1, where $\Delta$$_{CEF}$ labels the size of the
crystal
field splitting.  In general, equations~\ref{eq:71} -~\ref{eq:8-2} can be
expressed as
\begin{equation}  \label{eq:newstates}
|\Gamma,\alpha\rangle=\sum_{m}c_{\Gamma\alpha m}|m\rangle,
\end{equation}
where the coefficients $c_{\Gamma\alpha m}$ can be
read directly from the equations.

We can use equation~\ref{eq:newstates} to construct the
matrix element, V$_{\Gamma\alpha\sigma}$($\vec k$), for hybridization between
a crystal field state with quantum numbers $\Gamma$, $\alpha$ and a plane wave
conduction state
with crystal momentum $\vec k$ and spin $\sigma$.  In the case of a single ion
in the full
J=5/2 manifold, one may use the form derived by Coqblin
and Schrieffer \cite{Coqblin 69},
\begin{equation}  \label{eq:coqsch}
V_{m\sigma}(\vec k)=-\sqrt{\frac{4\pi}{3}}(-i)^3\sigma
V_{ok}\sqrt{\frac{7-2m\sigma}{14}}
Y^{*}_{3,m-\frac{\sigma}{2}}(\hat k),
\end{equation}
(-$5/2\leq m\leq 5/2$),
$\sigma=\pm$1 is the (pseudo)spin index, and the angular dependence is
in the spherical harmonic.  $V_{ok}$ denotes the dependence of the
hybridization strength on the
{\em magnitude} of the momentum, which will be important only near the zone
center ({\em i.e.}
near $|k|$=0).  In fact, we can write $V_{ok}=V_o g(k)$, where $g(k)$ is a
function of $|\vec k|$ that
goes to zero at the zone center like $k^3$.  $V_o$ is the bare hybridization
strength.
Because the crystal field split $4f$ states
are just linear combinations of the $J=5/2$ states, the hybridization matrix
elements in cubic symmetry are
\begin{equation} \label{eq:crystalv}
V_{\Gamma \alpha \sigma}(\vec k)=\sum_{m=-5/2}^{5/2}c_{\Gamma \alpha
m}V_{m\sigma}(\vec k).
\end{equation}
Figure 2 represents the hybridization process pictorially.


\begin{center}
{\Large III. HAMILTONIAN}
\end{center}
\vspace{0.4cm}

We start with a few words on the important energy scales of \cc,
based on the discussion of Kang
{\em et al.}\cite{Kang 90} on the electron spectroscopic data available as of
1990.
They analyzed their
own data of the Ce $3d$ x-ray photoelectron spectrum (XPS) and $4f$
bremsstrahlung isochromat
spectrum (BIS); they also analyzed Ce $4f$ resonant photoelectron data (RESPES)
of
Parks {\em et al.}\cite{Parks 84}.
Kang and co-workers calculated the appropriate one-electron spectra from the
impurity Anderson model, which showed reasonably
good agreement with the data.  On the basis of such a calculation,
the authors claim that Coulomb energy for double occupation of a Ce $4f$ site
is $U\approx$ 7$eV$.

The large Coulomb energy for \cc prompts us to take the limit in which $U$ goes
to infinity,
thereby forbidding hybridization processes that give rise to $4f^1\rightarrow$
$4f^2$ valence
fluctuations.  This is a technical simplification for us, but even though $U$
is indeed very
large by solid-state physics standards, it may be that the physics of finite
$U$ is crucial to the
understanding of heavy Fermions.  Recently Dan Cox\cite{Cox Quad} has proposed
the
Quadrupolar Kondo Model (or Two-Channel Kondo Model)
as an explanation of the superconductivity and of the possible non-Fermi liquid
behavior in
uranium based heavy Fermions.  The physics needed to get non-Fermi liquid
behavior
derives from the existence of enough independent channels of conduction
electrons to
overcompensate the effective $4f$ ``spin''.
A finite-$U$ version of the Anderson model, as applied to a Ce {\em impurity},
can
exhibit a two-channel Kondo effect.  The crystal field split multiplet
structure here is the key.
Taking the $\Gamma_7$ doublet as the low-lying
multiplet for the $4f^1$ configuration, and a $\Gamma_3$ ({\em non-magnetic})
doublet as the low
lying multiplet for the $4f^2$ configuration (as is appropriate for cubic
symmetry), a two-channel
Kondo effect is possible. Conduction states of $\Gamma_8$ symmetry can mix the
$4f^1$ and $4f^2$
configurations\cite{coxprivate}, and conduction states of $\Gamma_7$ symmetry
can mix the
$4f^1$ and $4f^0$ configurations. If the hybridization of $\Gamma_8$ conduction
states
is stronger than hybridization of $\Gamma_7$ conduction states,
the physics of the two-channel Kondo effect will determine the low temperature
properties of the model.
It may be that the finite-U Anderson model is a better starting point for
heavy fermion systems.  Nevertheless, the infinite $U$ limit is a reasonable
simplification (at least for \cc)
of an already difficult problem and warrants study in its own right.

Our Hamiltonian can be written as a combination of terms:
$H=H_c+H_f+H_{mix}+H_{constraint}$.
The kinetic energy of the conduction electrons is given by,
\begin{equation}  \label{eq:Hc}
H_c=\sum_{\vec k \sigma}\xi_{\vec k}c^{\dagger}_{\vec k\sigma}c_{\vec k\sigma},
\end{equation}
where
\begin{equation}  \label{eq:free}
\xi_{\vec k}=\frac{\hbar^2k^2}{2m}-\mu_o,
\end{equation}
is the plane-wave dispersion.  The zero of energy for this calculation will be
taken with respect
to $\mu_o$, the chemical potential of the conduction electrons in the absence
of hybridization.

The $4f$ electron site energy is
\begin{equation}  \label{eq:Hf}
H_f=\sum_{\vec R\ga}E_{\Gamma}f^{\dagger}_{\vec R\ga}f_{\vec R\ga},
\end{equation}
where $\vec R$ is the site index in real-space, and $\ga$ are the crystal field
quantum numbers.
The operator $f_{\vec R\ga}$ destroys a $4f^1$ configuration at lattice site
$\vec R$, in which
the crystal field state $\ga$ is initially occupied.
{}From the photoemission data\cite{Kang 90}, we take the energy of the
$\Gamma_7$ doublet to be -2.0 eV, {\em i.e.}
$E_7=$-2.0 eV.  From the inelastic neutron scattering data\cite{Horn 81} and
the high temperature
specific heat\cite{Bredl 85}, we take the $\Gamma_8$ level to lie 360 meV above
the $\Gamma_7$ level,
\begin{equation}  \label{eq:E8}
E_8=E_7+\Delta_{CEF}=-1.964\hspace{0.2cm}{\em eV}.
\end{equation}

The hybridization, or mixing, term is
\begin{equation}  \label{eq:Hmix}
H_{mix}=\frac{1}{\sqrt{N_s}}\sum_{\vec k \sigma\vec
R\Gamma\alpha}\biggl[V_{\Gamma\alpha\sigma}
(\vec k)c^\dagger_{\vec k\sigma}f_{\vec R\Gamma\alpha}b^{\dagger}_{\vec R}
e^{i\vec k \cdot \vec R}
 + H.c.\biggr].
\end{equation}
The operator $b^{\dagger}_{\vec R}$ is a slave Boson creation operator, which
creates a $4f^0$
configuration, or hole, at lattice site $\vec R$.  $N_s$ is the number of
lattice sites.  This
combination of conduction, $4f$, and
slave Boson operators was first applied to the Anderson model by
Barnes\cite{Barnes 76}, and was
later reintroduced by Coleman\cite{Coleman 84}.  $H_{mix}$ contains only
Bosonic or Fermionic operators, and so Wick's theorem is applicable.  Use of
the more cumbersome
Hubbard projection operators,
\begin{equation}  \label{eq:Hubbard}
X_{0\ga}=|0\rangle\langle \ga|,
\end{equation}
would make Feynman diagrammatic procedures invalid, since the
Hubbard operators do not obey standard commutation relations.
The cubic symmetry is reflected in the structure of the anisotropic function
V$_{\Gamma\alpha\sigma}$($\vec k$), which has a significantly different $\vec
k$ dependence than the
Coqblin-Schrieffer form V$_{m\sigma}$($\vec k$).  This new anisotropy will
affect
quasiparticle interactions differently than in the case of spherical symmetry.

Finally, there is the constraint term, which is introduced with a Lagrange
multiplier,
$i\lambda_{\vec R}$, insuring that the {\em total} occupancy (Fermions plus
Bosons) at the
Ce sites is unity,
\begin{equation}  \label{eq:Hconstraint}
H_{constraint}=\sum_{\vec R}i\lambda_{\vec R}\biggl(f^{\dagger}_{\vec
R\ga}f_{\vec R\ga}
+b^{\dagger}_{\vec R}b_{\vec R}-Q\biggr).
\end{equation}
Note that $Q$=1 is the physically meaningful value in this case.

There are a few technical points we need to mention here.  As discussed by Read
and Newns
\cite{Read 83a} in the functional integral approach to the
infinite-U Anderson impurity, it is useful to write the slave Boson operator as
the
product of a modulus and a phase factor, which in the lattice problem picks up
a site index, $\vec R$,
\begin{equation}  \label{eq:b}
b_{\vec R}=s_{\vec R}e^{i\theta_{\vec R}}.
\end{equation}
A gauge transformation for the $f$ operators,
$$f_{\vec R\ga}\rightarrow f_{\vec R\ga}e^{-i\theta_{\vec R}},$$
then absorbs all the phase factors.  In the functional integral approach,
furthermore, the Lagrangian
corresponding to our Hamiltonian
(equations~\ref{eq:Hc},~\ref{eq:Hf},~\ref{eq:Hmix}, and
\ref{eq:Hconstraint})
contains an (imaginary) time derivative of the slave Bosons,
$$b^{*}_{\vec R}(\tau)\frac{\partial b_{\vec R}(\tau)}{\partial \tau},$$
which, from equation~\ref{eq:b}, introduces a factor of the phase velocity,
$\dot\theta_{\vec R}$,
into the Lagrangian.  Traditionally, this phase velocity is absorbed into the
Lagrange multiplier, elevating
it to the status of a dynamical field, $i\lambda_{\vec R}(\tau)$.  In this
paper, since we have
used the Hamiltonian formalism, we remark that, in the limit of static slave
Bosons, the
Lagrange multiplier plays the same role as it does in the functional integral
formalism\cite{Millis 86}.
We have a 2$\times$2 matrix Green function for the slave Boson and Lagrange
multiplier,
which will have elements composed of averages over the following combinations
of fields: $ss$, $s\lambda$,
and $\lambda\lambda$.

Our calculation is based on previous $1/N$ calculations of
the type applied to the $SU(N)$ model, where $N$ is the degeneracy of the $4f$
multiplet.  In the $SU(N)$
case, in
order to have a well defined Kondo temperature in the limit of large $N$, it
was necessary to assume
the bare hybridization strength, $V_o$, scaled like $1/\sqrt{N}$\cite{Millis
86}.  We do the same here by
defining a rescaled hybridization matrix element,
\begin{equation}  \label{eq:scaledV}
\tilde V_{\ga\sigma}(\vec k)\equiv \sqrt{N_{\Gamma}}V_{\ga\sigma}(\vec k),
\end{equation}
where we assume that $\tilde V_{\ga\sigma}(\vec k)$ is of order 1.  Note that
$N_{\Gamma}$ is the degeneracy of the $\Gamma$ crystal field multiplet.  Also,
because the
hybridization matrix element and the Boson operator, $s_{\vec R}$, appear
together in $H_{mix}$, we
define a scaled Boson operator,
\begin{equation}  \label{eq:scaleds}
\tilde s_{\vec R\Gamma}\equiv \frac{s_{\vec R}}{\sqrt{N_\Gamma}},
\end{equation}
where we assume that $\tilde s_{\vec R\Gamma}$ is also of order 1.

\begin{center}
{\Large IV. MEAN FIELD APPROXIMATION}
\end{center}
\vspace{0.4cm}

In this section, we discuss the properties of our Hamiltonian at mean field
level, where
we assume both the Bose operator and the Lagrange multiplier are uniform in
space.   In this limit,
the Hamiltonian in k-space takes the following form:
$$H_{MF}=\sum_{\vec k\sigma}\xi_{\vec k}c^{\dagger}_{\vec k\sigma}c_{\vec
k\sigma}+\sum_{\vec k\Gamma\alpha}
\epsilon_{\Gamma}f^{\dagger}_{\vec k\Gamma\alpha}f_{\vec k\Gamma\alpha}$$
$$+\sum_{\vec k\sigma\Gamma\alpha}\biggl[\tilde s_{o\Gamma}\tilde
V_{\Gamma\alpha\sigma}(\vec k)
c^{\dagger}_{\vec k\sigma}f_{\vec k\Gamma\alpha}+H.c.\biggr]$$
\begin{equation} \label{eq:hmf}
+\frac{N_s}{2}\sum_{\Gamma}N_{\Gamma}i\lambda_o\biggl(\tilde
s^{2}_{o\Gamma}-q_{o\Gamma}\biggr).
\end{equation}
where $\epsilon_{\Gamma}\equiv E_{\Gamma}+i\lambda_o$ is the shifted energy of
the $\Gamma$ multiplet.
We assume that $q_{o\Gamma}$ is of order 1, but technically, when it comes down
to getting
numerical results, we know that $q_{o\Gamma}=1/N_{\Gamma}$.
Note that $\tilde s_{o\Gamma}$ and $\lambda_o$
are the mean field values of the Bose operator and Lagrange multiplier,
respectively ($\tilde s_{o\Gamma}
=s_o/\sqrt{N_{\Gamma}}$).

It is straightforward to diagonalize $H_{MF}$ and obtain the quasiparticle
states and energies.
Because the $\Gamma_8$ states are four-fold degenerate, there is one
non-hybridizing
quasiparticle band
of $\Gamma_8$ symmetry.  The secular equation which gives the quasiparticle
energies $E_{n\vec k}$
($n$=1,2,3 is the band index) is
\begin{equation} \label{eq:fullsecular}
\bigl(\epsilon_8-E_{n\vec k}\bigr)\biggl[\bigl(\xi_{\vec k}-E_{n\vec
k}\bigr)-\sum_{\alpha_7}\frac{\tilde s^{2}_{o7}
|\tilde V_{7\alpha\sigma}|^2}{\bigl(\epsilon_7-E_{n\vec
k}\bigr)}-\sum_{\alpha_8}\frac{\tilde s^{2}_{o8}
|\tilde V_{8\alpha\sigma}|^2}{\bigl(\epsilon_8-E_{n\vec k}\bigr)}\biggr]=0,
\end{equation}
where the sum $\sum_{\alpha_7(8)}$ means only the states in the $\Gamma_{7(8)}$
doublet are summed.
Although technically it is possible to find analytic solutions
of equation~\ref{eq:fullsecular}, the analytic expressions are too cumbersome
to be useful;
thus we calculated the
roots numerically.  The three quasiparticles bands are plotted in figure 3
along two different
directions in the cubic Brillouin zone.  Figure 3 shows that along the axes of
the zone, for example along
the $\Gamma X$ direction, the states of $\Gamma_7$ symmetry can not hybridize
with the conduction states.
This means that the matrix elements $V_{7\alpha\sigma}(\vec k)$ vanish along
these special directions and that
there is no gap between the first and second quasiparticle bands (see figure
3(b)).  Such behavior
has been discussed by Martin\cite{Martin 82}.

The quasiparticle states can, quite generally, be written as a combination of
plane wave and
crystal field states,
\begin{equation} \label{eq:eigvector}
|Q_{\vec k n\sigma}\rangle=A_n(\vec k)\biggl[|\vec k\sigma\rangle
-\sum_{\ga}\frac{\tilde s_{o
\Gamma}\tilde V^{*}_{\ga\sigma}(\vec k)}{\epsilon_{\Gamma}-E_{n\vec
k}}|\ga\rangle\biggr],
\end{equation}
where $|\vec k\sigma\rangle$ is a plane wave state.
The {\em anisotropic} normalization function is
\begin{equation} \label{eq:fulla}
A^{2}_n(\vec k)=\biggl[1+\frac{1}{2}\sum_{\ga\sigma}\frac{\tilde
s^{2}_{o\Gamma}|\tilde V_{\ga\sigma}(\vec k)|^2}
{\bigl(\epsilon_{\Gamma}-E_{n\vec k}\bigr)^2}\biggr]^{-1}.
\end{equation}
%
For given crystal field quantum numbers, $\ga$, it
is possible to sum over the pseudo-spin variable $\sigma$ in
equation~\ref{eq:fulla}.  For convenience,
we define a function
\begin{equation}  \label{eq:mu}
\mu_{\ga\Gamma^\prime\alpha^\prime}(\vec k)\equiv\sum_{\sigma}\tilde
V^{*}_{\ga\sigma}(\vec k)\tilde V_{\sigma
\Gamma^{\prime}\alpha^{\prime}}(\vec k).
\end{equation}
Then the sum over the pseudospin variable in equation~\ref{eq:fulla} is just
the diagonal element
$\mu_{\ga\ga}(\vec k)$.  All possible (non-zero) forms for
the function $\mu_{\ga\Gamma^\prime\alpha^\prime}(\vec k)$ are shown in Table
1, where the dependence on the
{\em magnitude} of $\vec k$ has been divided out.  Using the results of this
table, it is possible to arrive
at the following expression for the normalization function, which is valid at
any point in the Brillouin zone:
\begin{equation}  \label{eq:simplea}
A^{2}_{1}=\frac{\biggl(\frac{T_{o7}}{s_oV_o}\biggr)^2}{\frac{1}{3}+\frac{2}{3}\biggl(\frac
{T_{o7}}{T_{o8}}\biggr)^2-\frac{2\sqrt{\pi}}{9}\biggl[1-\biggl(\frac{T_{o7}}{T_{o8}}\biggr)^2
\biggr]\biggl[Y_{40}(\hat k)+\sqrt{\frac{5}{14}}\biggl(Y_{44}(\hat
k)+Y_{4-4}(\hat k)(\biggr)\biggr]}.
\end{equation}
Figure 4 shows a plot of equation~\ref{eq:simplea} along the equator of a
spherical Fermi surface, with $\phi$ denoting the azimuthal angle measured with
respect to
a coordinate axis.  The extremely sharp variations near the axes represent the
vanishing of the $\Gamma_7$ hybridization matrix elements.
We call the six points where the Brillouin zone axes intersect
the Fermi surface ``hot-spots''; they must be handled with care when averaging
the
quasiparticle scattering amplitude over the Fermi surface.

We would like to remark that, in this model, the existence of ``hot-spots'' is
a manifestation
of the lowering of the symmetry
below spherical.  In spherical symmetry, there is a sum rule for the matrix
elements that renders the
normalization function, $A^{2}$, isotropic in k-space\cite{Lavagna
87},\cite{Zhang 87}.  In our case, the
matrix elements can not be simplified to an isotropic function.  This means
that ``hot-spots'' can occur anywhere
in the Brillouin zone where some subset of the hybridization matrix elements,
$\tilde V_{\ga\sigma}(\vec k)$,
vanishes.  This is a rather general statement that relies only on the symmetry
being lower than spherical
and should not be unique to cubic lattices.

We can relate the fermionic creation and destruction operators in the
original basis (the $c$ and $f$ operators) to creation and destruction
operators in the quasiparticle
basis, $Q^{\dagger}_{\vec k n\sigma}$, $Q_{\vec k n\sigma}$.  We find that
\begin{equation} \label{eq:candq}
c_{\vec k\sigma}=\sum_n A_n(\vec k)Q_{\vec k n\sigma},
\end{equation}
\begin{equation} \label{eq:fandq}
f_{\vec k\ga}=-\sum_{n\sigma}\frac{A_n(\vec k)\tilde s_{o\Gamma}\tilde
V^{*}_{\ga\sigma}(\vec k)}
{\epsilon_{\Gamma}-E_{n\vec k}}Q_{\vec k n\sigma}.
\end{equation}
These expressions are useful in constructing the two-quasiparticle scattering
amplitude.

In the last part of this section, we discuss the self-consistency equations
that arise when one demands
that the free energy in the mean field approximation be an extremum with
respect to the Bose
fields $s_o$ and $i\lambda_o$.
Since we have diagonalized the mean field Hamiltonian, equation~\ref{eq:hmf},
we can write it in terms of
the quasiparticle energies and operators:
\begin{equation}  \label{eq:quasihmf}
H_{MF}=\sum_{\vec k n\sigma}E_{n\vec k}Q^{\dagger}_{\vec k n\sigma}Q_{\vec k
n\sigma}+
\frac{N_s}{2}\sum_{\Gamma}N_{\Gamma}i\lambda_o\biggl(\tilde
s^{2}_{o\Gamma}-q_{o\Gamma}\biggr).
\end{equation}
The corresponding mean field free energy has the form
\begin{equation}  \label{eq:meanfree}
F_{MF}=\frac{N_s}{2}\sum_{\Gamma}N_{\Gamma}i\lambda_o\biggl(\tilde
s^{2}_{o\Gamma}-q_{o\Gamma}\biggr)
-\frac{1}{\beta}\sum_{\vec k n\sigma}\ln\bigl(1+e^{-\beta E_{n\vec k}}\bigr),
\end{equation}
where $\beta$ is the inverse temperature.
Requiring that $\partial F_{MF}/\partial i\lambda_o =$0 and $\partial
F_{MF}/\partial s_o =$0 yields the
equations
\begin{equation} \label{eq:dfdl}
\frac{1}{2}\sum_{\Gamma}N_{\Gamma}\bigl(\tilde
s^{2}_{o\Gamma}-q_{o\Gamma}\bigr)+
\frac{1}{N_s}\sum_{\vec k\sigma n}f(E_{n\vec k})\frac{\partial E_{n\vec
k}}{\partial i
\lambda_o}=0,
\end{equation}
and
\begin{equation} \label{eq:dfds}
2i\lambda_os_o+\frac{1}{N_s}\sum_{\vec k\sigma n}f(E_{n\vec k})\frac{\partial
E_{n\vec k}}
{\partial s_o}=0.
\end{equation}
In equations~\ref{eq:dfdl} and~\ref{eq:dfds} $f(E_{n\vec k})$ is the Fermi
function
evaluated at the quasiparticle energy $E_{n\vec k}$.
We also need an equation to fix the chemical potential of the {\em
quasiparticles},
$\mu$, which depends on the total number of electrons (conduction electrons,
$n_c$,
and f electrons, $n_f$) per unit cell
\begin{equation} \label{eq:chempot}
n_{total}=n_c+n_f=\frac{1}{N_s}\sum_{\vec k\sigma n}f(E_{n\vec k}).
\end{equation}

These three coupled integral equations, when solved self-consistently, give the
shifted 4f multiplet energies,
$\epsilon_{\Gamma}$, the value of the Bose field, $s_o$, and the quasiparticle
chemical potential,
$\mu$.  The input parameters are $n_{total}$, the quasiparticle filling factor,
the bare hybridization
strength, $V_o$, and the conduction electron filling factor, $n_c$.  The zero
of energy is always measured
relative the the chemical potential of the {\em conduction} electrons, $\mu_o$.
For $n_{total}=$2, the
lowest quasiparticle band is completely filled, and the system is a Kondo
insulator.  We have consistently used $n_{total}=$1.5, which insures that we
have a metal.

To get numerical self-consistent solutions, we found it necessary to write
these
equations in terms of energy integrals with the appropriate density of states,
\begin{equation} \label{eq:dos}
\frac{1}{N_s}\sum_{\vec k\sigma}\rightarrow 2\int d\xi N(\xi)\int
\frac{d\Omega}{4\pi},
\end{equation}
where $N(\xi)$ is the density of states per spin for the unhybridized
conduction electrons and d$\Omega$ is an element of solid angle.
Note that
in the $SU(N)$ model described previously, the spin degeneracy would contribute
a prefactor of $N$ (instead of 2) in equation~\ref{eq:dos}. For free electrons
in three dimensions
the density of states is proportional to the square root of $\xi$.

To proceed, we make one approximation.  We assume that surfaces of constant
energy for
the quasiparticle states are spherically symmetric.  Near the zone center, this
is
exactly correct, and there is no approximation at all.
Near the zone boundary, the equal-energy surfaces become
distorted from spheres due to the constraints of $\Gamma_7$ symmetry.
It is important to note, however, that in
equations~\ref{eq:dfdl},~\ref{eq:dfds}, and~\ref{eq:chempot}, the {\em
strongest}
angular dependence comes from the anisotropic matrix elements, which we treat
{\em exactly}.
That is, we believe
the angular dependence of the quasiparticle bands is not as important as that
of the hybridization
matrix elements.  For example, near an axis of the Brillouin zone, the
$\Gamma_7$ matrix
elements are going to zero.  So, even if the
quasiparticle energies surfaces are distorted from spheres near the axes, the
sensitivity
of the self-consistency equations to this distortion would be lessened by the
presence of the small
$V_{7\alpha\sigma}$ terms in the numerator.  Thus, it should be a reasonable
approximation
to treat the mixing matrix elements as having all the angular dependence.

Using the secular equation for the quasiparticle band energies,
equation~\ref{eq:fullsecular}, we can
calculate all the necessary derivatives of the quasiparticle energies found
in equations~\ref{eq:dfdl} and~\ref{eq:dfds}.  After averaging
the anisotropic matrix elements over the Fermi surface, we are left with the
following three equations
to be solved self-consistently:
\begin{equation} \label{eq:finaldfdl}
\frac{1}{2}\sum_{\Gamma}N_{\Gamma}\bigl(\tilde
s^{2}_{o\Gamma}-q_{o\Gamma}\bigr)+
2\frac{s^{2}_{o}V^{2}_{ko}}{3}\int_{-D}^{\mu}dE
\hspace{0.2cm}N\bigl(\xi(E)\bigr)\biggl[\frac{1}
{\bigl(\epsilon_7-E\bigr)^2}+\frac{5}{\bigl(\epsilon_8-E\bigr)^2}\biggr]=0,
\end{equation}
\begin{equation} \label{eq:finaldfds}
2i\lambda_o-2\frac{V^{2}_{ko}}{3}\int_{-D}^{\mu}dE
\hspace{0.2cm}N\bigl(\xi(E)\bigr)\biggl[\frac{1}
{\epsilon_7-E}+\frac{5}{\epsilon_8-E}\biggr]=0,\hspace{0.2cm}{\rm and}
\end{equation}
\begin{equation} \label{eq:finaltotal}
n_{total}-2\int_{-D}^{\mu}dE \hspace{0.2cm}N\bigl(\xi(E)\bigr)\biggl[
1+\frac{1}{3}\frac{s^{2}_{o}V^{2}_{ok}}{\bigl(\epsilon_7-E\bigr)^2}
+\frac{5}{3}\frac{s^{2}_{o}V^{2}_{ok}}{\bigl(\epsilon_8-E\bigr)^2}\biggr],
\end{equation}
where $-D$ is the energy at the bottom of the lowest quasiparticle band.

We have taken the limit of zero temperature to arrive at
equations~\ref{eq:finaldfdl}-\ref{eq:finaltotal}.
As a consequence, only the first quasiparticle band, $E_1$, contributes at mean
field; for simplicity, we have
dropped the band subscript ``1''.
We consider two different sets of solutions, corresponding to the {\em
conduction} electron filling factors of
$n_c=$0.5 (which we call set (a)) and $n_c=$0.8 (which we call set (b)).
We define a Kondo temperature in the lattice for both
$\Gamma_7$ and $\Gamma_8$ multiplets by
\begin{equation}  \label{eq:Tk}
T_{o\Gamma}\equiv\epsilon_{\Gamma}-\mu.
\end{equation}
The reader is reminded that
$\mu$ is the {\em quasiparticle} chemical potential.  The motivation for
defining the Kondo temperature as
the difference between the shifted multiplet energy ($\epsilon_{\Gamma}$) and
the quasiparticle chemical
potential, is that in the $SU(N)$ model this difference has exactly the same
structure as the Kondo temperature for
the impurity problem.  That is, in the $SU(N)$ model, one finds
that\cite{Fulde}
\begin{equation}  \label{eq:suntk}
\epsilon -\mu=De^{-|E_f|/NN(0)V^{2}_{o}},
\end{equation}
where $D$ is the half bandwidth, $N(0)$ is the (assumed flat) conduction
electron density of states,
$E_f$ is the {\em unshifted} $4f$ multiplet energy,
and $V_o$ is the bare hybridization strength.  In table 2, we see that both
parameter sets (a) and (b) have
approximate Kondo temperatures (for the $\Gamma_7$ doublet) of 10 K, which,
based on the neutron scattering
quasielastic linewidth\cite{Horn 81}, is a reasonable estimate for \cc.
Note, also, that the Kondo temperature for the $\Gamma_8$ quartet is dominated
by the crystal field splitting,
\begin{equation}  \label{eq:To8}
T_{o8}=T_{o7}+\Delta_{CEF}\approx 370\hspace{0.2cm}K,
\end{equation}
where $T_{o7}/T_{o8}\approx$0.027.
Figure 5 plots the lowest quasiparticle bands for both mean field parameter
sets (a) and (b).
The top of the first quasiparticle
band is just below the shifted $\Gamma_7$ energy, $\epsilon_7$.  The {\em
quasiparticle} chemical potential cuts
through the flat part of the first band in both cases, giving rise to a very
large quasiparticle density of
states at the Fermi surface.

\begin{center}
{\Large V. EFFECTIVE MAGNETIC MOMENT}
\end{center}
\vspace{0.4cm}

In this section we discuss our results for the effective magnetic moment of
the quasiparticle states when averaged over the Fermi surface.  This
calculation
was motivated by the work of Zou and Anderson\cite{Zou 86}, who wished to
explain
how the small Wilson ratios, $R$, seen in heavy Fermion superconductors (for
\cc, $R\approx$0.5\cite{Stewart 84})
could be reconciled with the presence of strong spin fluctuations.
If spin fluctuations are indeed the source of the superconductivity in heavy
fermion compounds\cite{Anderson 84},
one would expect a Wilson ratio greater than unity.
Zou and Anderson have claimed that, even in the presence of strong spin
fluctuations, the Wilson ratio could be
reduced due to the anisotropic hybridization between the conduction and the $f$
electrons.
They calculated the quasiparticle states, including spin-orbit coupling,
from the relativistic KKR equation. Their states are identical
to those found via a mean-field approximation to the
infinite-U Anderson lattice\cite{Millis 86}, \cite{Tesanovic 86},
\cite{Zhang 86}, or via
Gutzwiller projection techniques\cite{Rice 85}.

In spherical symmetry, for a total angular momentum of
$J=5/2$, the quasiparticle states are
\begin{equation}  \label{eq:states}
|Q_{\vec k n\sigma}\rangle=A_n(|\vec k|)\biggl[|\vec
k\sigma\rangle-\sum_{m=-5/2}^{5/2}\frac{s_oV_{m\sigma}(\vec k)}
{(\epsilon-E_{nk})}|m\rangle\biggr],
\end{equation}
where the normalization for the lowest energy band is
\begin{equation}  \label{eq:normalization}
A^{2}_{1}(|\vec
k|)=\biggl[1+\frac{s^{2}_{o}V^{2}_{o}}{(\epsilon-E_{1k})^2}\biggr].
\end{equation}
In the preceding equations, $E_{nk}$ is the quasiparticle energy for band $n$
and
$\epsilon$ is the energy of the shifted $4f$ states.  Note that the
normalization
function is {\em isotropic} and depends only on the magnitude of the
wavevector.
For the case of $J=5/2$ there is a four-fold degenerate band of energy
$\epsilon$, since
two of the six total
$f$ states have hybridized with the conduction states.

Zou and Anderson calculated the magnetic moment of the quasiparticle states in
the lowest band, which
we denote as
\begin{equation}  \label{eq:moment}
\langle Q_{\vec k 1\sigma}|\hat\mu_z|Q_{\vec k
1\sigma}\rangle\equiv\mu^{\sigma\sigma^\prime}_{z}(\vec k),
\end{equation}
where the magnetic moment operator is
\begin{equation}  \label{eq:operator}
\hat\mu_z=\hat l_z +2\hat s_z=g_J\hat J_z,
\end{equation}
$g_J$ being the g-factor for the total angular momentum, $J$.
The effective magnetic moment is the average of $\mu^{\sigma\sigma^\prime}_{z}$
over the spherical
Fermi surface,
\begin{equation}  \label{eq:averagemoment}
\mu^{2}_{eff}=\int\frac{d\hat k}{4\pi}\biggl[\mu^{+1+1}_{z}(\vec
k)^2+\mu^{+1-1}_{z}(\vec k)^2\biggr],
\end{equation}
and the result is $\mu^{2}_{eff}$=1.16$\mu^{2}_{B}$, where $\mu_{B}$ is the
Bohr-magneton. Note that the bare
moment for a $f^1$ state is $\mu$=2.54$\mu_B$, so that hybridization has
reduced the size of the
moment.

The Pauli susceptibility for the quasiparticles is a function of the effective
magnetic moment
and the renormalized density of states at the Fermi surface, $\tilde N(0)$
($\tilde N(0)=\frac{m^*}{m}N(0)$, where $m^*$ is the quasiparticle effective
mass),
\begin{equation}  \label{eq:pauli}
\chi_{Pauli}=2\mu^{2}_{eff}\tilde N(0).
\end{equation}
Substituting this value for the Pauli susceptibility into the expression for
the Wilson ratio gives
\begin{equation}  \label{eq:reducedR}
R_{reduced}=\biggl(\frac{\mu_{eff}}{\mu}\biggr)^2\frac{1}{1+F^{a}_{0}}=\frac{0.18}{1+F^{a}_{0}}.
\end{equation}
In this way, it would be possible to have a large Stoner factor,
$(1+F^{a}_{0})^{-1}$, as befits the
presence of strong spin fluctuations, but the reduced magnetic moment could
still account for a small
experimentally observed Wilson ratio.

Cox\cite{Cox 87},Zhang and Lee\cite{Zhang 87a}, and Aeppli and Varma\cite{Varma
87}
pointed out the need to
include the Van Vleck susceptibility in the calculation.
The value for $\chi$ in equation~\ref{eq:reducedR} must be the total
susceptibility, as seen by experiment. In the case of the uniform
susceptibility, the Van Vleck
contribution represents direct transitions from the Fermi energy in the lowest
band to higher energy bands.
The dominant contribution must come from transitions to the four-fold
degenerate band
lying at an energy $T_o$ above the Fermi energy, where $T_o$ is the Kondo
temperature
in spherical symmetry.
The Van Vleck susceptibility then has the structure
\begin{equation}  \label{eq:vanvleck}
\chi_{VV}(\vec q =0)=2\mu^{2}_{B}\sum_m\int\frac{d\hat k}{4\pi}\frac{|\langle
Q_{\vec k 1\sigma}|\hat\mu_z|
m\rangle|^2}{T_o},
\end{equation}
where the quasiparticle density of states $\tilde N(0)\approx 1/T_o$.  The
small energy denominator
(for \cc, $T_o\approx$ 1 meV)
makes this interband contribution to the susceptibility comparable to the
contribution from
the Pauli susceptibility. Cox and Zhang and Lee showed that the total
susceptibility was given by
\begin{equation}  \label{eq:netchi}
\chi_{total}=\chi_{Pauli}+\chi_{VV}=\frac{1}{3}g^{2}_{J}J(J+1)\mu_{B}^{2}[2\tilde N(0)],
\end{equation}
which depends on the {\em bare} magnetic moment.  Thus the inclusion of the Van
Vleck
susceptibility would alter Zou and Anderson's argument.

We now wish to discuss what happens when crystal electric fields of cubic
symmetry are included.  The quasiparticle states are given by
equation~\ref{eq:eigvector}, where the local crystal field states are
orthonormal to each other.
Note that in this case the dispersionless band lies at an energy $\Delta_{CEF}$
above
the Fermi energy, where the crystal field splitting is much larger than the
Kondo temperature.
In order to calculate the effective magnetic moment, we need the expectation
value of equation~\ref{eq:moment}
for the lowest quasiparticle band.  We find that we can write
equation~\ref{eq:moment} in the form
\begin{equation}  \label{eq:mess}
\mu^{\sigma\sigma^\prime}_{z}(\vec k)=A^{2}_{1}(\vec
k)\biggl[\sigma\delta_{\sigma\sigma^\prime}+
\sum_{\gp\ap\ga m}\frac{mg_Jc_{\gp\ap m}s_{o\gp}V^{*}_{\gp\ap\sigma}(\vec
k)V_{\ga\sigma^\prime}(\vec k)
s_{o\Gamma}c_{\ga m}}{(\epsilon_{\gp}-E_{1\vec k})(\epsilon_{\Gamma}-E_{1\vec
k})}\biggr].
\end{equation}
Using equation~\ref{eq:averagemoment}, we find the average moment is
\begin{equation}  \label{eq:averagecubicmoment}
\mu^{2}_{cubic,eff}=0.583\mu^{2}_{B},
\end{equation}
for mean-field parameter set (a).
For a free $\Gamma_7$ doublet the average moment is
\begin{equation}  \label{eq:g7}
\mu^{2}_{7}=\frac{25}{49}\mu^{2}_{B}.
\end{equation}
The effective magnetic moment in cubic symmetry is slightly larger than the
value for a free $\Gamma_7$ moment.  This was expected by Cox\cite{Cox 87},
who predicted the effective moment would have the structure
\begin{equation}  \label{eq:magmoment}
\mu^{2}_{eff}=\mu^{2}_{7}\biggl[1+\alpha\sqrt{\frac{m}{m^{*}}}\biggr],
\end{equation}
where $\alpha$ is a prefactor that could be as big as about 10. The
contributions of order $\sqrt{m/m^*}$ come from the six regions on
the Fermi surface which intersect the axes of the cubic Brillouin zone.
These are the so-called ``hot-spots'', which we have already mentioned.
At these points on the Fermi surface,
the plane wave conduction states and the localized $\Gamma_7$ states
can not hybridize.  Thus we expect the $g$-factor at these "spots" to
go back to the free electron value of 2, with the result that the contribution
to $\mu_{eff}$ from the ``hot-spots'' increases the effective moment
above that of a free moment of $\Gamma_7$ symmetry.
For parameter set (a), we have
$$\frac{m}{m^*}=\biggl(\frac{T_{o7}}{s_oV_o}\biggr)^2=1.33\times10^{-4}.$$
Equation~\ref{eq:magmoment} is applicable for the quoted values of
$\mu^{2}_{eff}$ and $\mu^{2}_{7}$ if $\alpha$=9.23, which is a reasonable
value.

Finally, we substitute $\mu^{2}_{eff}$ into the expression for the Wilson
ratio,
\begin{equation}  \label{eq:Wilson}
R=\biggl(\frac{\pi k_B}{\mu_{eff}}\biggr)^2\frac{\chi(0)}{\gamma(0)},
\end{equation}
where $\chi(0)$ is the low temperature susceptibility, $\gamma(0)$ is
the linear coefficient of specific heat, $k_B$ is Boltzmann's constant,
and $\mu_B$ is the Bohr magneton.  Using $\chi(0)$=0.019 emu/mole (for a
magnetic
field along the c axis of the tetragonal unit cell)\cite{Batlogg 84}, and
$\gamma(0)$=1000 mJ/mole-K$^2$ \cite{Steglich 79} gives $R=2.38$.
It is possible that experimentally quoted results of approximately 0.5 for the
Wilson ratio
\cite{Stewart 84} are too small since the bare effective moment of
$\mu^{2}_{eff}=$2.54$\mu^{2}_{B}$
was used in equation~\ref{eq:magmoment}.  We also find it interesting to note
that for the
two-channel Kondo impurity, the Wilson ratio is approximately 2.6.

\begin{center}
{\Large VI. HYBRIDIZATION DRESSED PROPAGATORS}
\end{center}
\vspace{0.4cm}

With the eigenstates of the mean field Hamiltonian written in the
undiagonalized basis, there are
three fermionic
Green functions:  G$_{\ga\gp\ap}$(the f Green function); G$_{\sigma}$(the
conduction
Green function); and G$_{\ga\sigma}$(the off-diagonal, or mixing, Green
function).  All
three are defined below in terms of Fock space operators:
\begin{equation} \label{eq:defgf}
G_{\ga\gp\ap}(\vec k,\tau)\equiv -\langle T_{\tau}f_{\vec
k\ga}(\tau)f^{\dagger}_{\vec k\gp\ap}
(0)\rangle,
\end{equation}
\begin{equation} \label{eq:defgc}
G_{\sigma}(\vec k,\tau)\equiv -\langle T_{\tau}c_{\vec
k\sigma}(\tau)c^{\dagger}_{\vec k\sigma}(0)
\rangle,
\end{equation}
\begin{equation} \label{eq:defgmix}
G_{\ga\sigma}(\vec k,\tau)\equiv -\langle T_{\tau}f_{\vec
k\ga}(\tau)c^{\dagger}_{\vec k\sigma}(0)
\rangle,
\end{equation}
where T$_{\tau}$ is the imaginary time ordering operator. It is easy to see,
diagrammatically,
how these hybridization dressed Green functions can be calculated.  Figures
6,7, and 8 show
the expansions for $G_{\ga\gp\ap}$, $G_{\sigma}$, and $G_{\ga\sigma}$,
respectively.
In terms of a complex frequency, $z$, the mean-field Green functions are:
\begin{equation} \label{eq:gff}
G_{\ga\gp\ap}(\vec
k,z)=\frac{1}{z-\epsilon_{\Gamma}}\biggl[\delta_{\Gamma\Gamma^\prime}
\delta_{\alpha\alpha^\prime}+\frac{1}{z-\epsilon_{\gp}}\frac{(z-\epsilon_7)(z-\epsilon_8)
\tilde s_{o\Gamma}\tilde V^{*}_{\ga\sigma}(\vec k)\tilde V_{\gp\ap\sigma}(\vec
k)
\tilde s_{o\gp}}{\bigl(z-E_{1\vec k}\bigr)\bigl(z-E_{2\vec
k}\bigr)\bigl(z-E_{3\vec k}\bigr)}
\biggr],
\end{equation}
\begin{equation} \label{eq:finalgcond}
G_{\sigma}(\vec k,z)=\frac{(z-\epsilon_7)(z-\epsilon_8)}{(z-E_{1\vec
k})(z-E_{2\vec k})(z-
E_{3\vec k})},
\end{equation}
\begin{equation} \label{eq:finalgmix}
G_{\ga\sigma}(\vec k,z)=\frac{\tilde s_{o\Gamma}\tilde V^{*}_{\ga\sigma}(\vec
k)\bigl(z-
\epsilon_{\Gamma^*}\bigr)}{\bigl(z-E_{1\vec k}\bigr)\bigl(z-E_{2\vec
k}\bigr)\bigl(z-E_{3
\vec k}\bigr)}.
\end{equation}
These propagators will be used to calculate the dressed {\em bosonic} Green
functions.

\begin{center}
{\Large VII. FLUCTUATIONS}
\end{center}
\vspace{0.4cm}

In this section, we shall calculate the Green functions for the Bosonic fields
in our Hamiltonian, namely $\tilde s_{\vec k\Gamma}$ and $i\lambda_{\vec k}$.
The Bosonic Green functions can be dressed (through the terms $H_{mix}$ and
$H_f$) by particle-hole excitations of the hybridized conduction and
$4f$ electron systems.  To proceed, we write the slave Boson and Lagrange
multiplier as follows:
\begin{equation}  \label{eq:sfluc}
\tilde s_{\vec k\Gamma}=\tilde s_{o\Gamma}\delta_{\vec k,0}+\delta\tilde
s_{\vec k\Gamma},
\end{equation}
\begin{equation}  \label{eq:lamfluc}
i\lambda_{\vec k}=i\lambda_o\delta_{\vec k,0}+i\delta\lambda_{\vec k},
\end{equation}
where $\delta\tilde s_{\vec k\Gamma}$ and $\delta\lambda_{\vec k}$ represent
fluctuations away from
the (self-consistent) mean field values.

It is easy to see how the particle-hole excitations dress the Bosonic Green
functions by
writing the {\em full} Hamiltonian in k-space,
$$ H=\sum_{\vec k\sigma}\xi_{\vec
k}c^{\dagger}_{\vec k\sigma}c_{\vec k\sigma} +
\sum_{\vec k\vec k^{\prime}\ga}f^{\dagger}_{\vec
k\ga}\biggl[E_{\Gamma}\delta_{\vec k
\vec k^\prime}+i\lambda_{\vec k-\vec k^\prime}\biggr]f_{\vec k^\prime\ga}$$
$$+\sum_{\vec k\vec k^\prime \ga\sigma}\biggl[\tilde
V_{\ga\sigma}(\vec k)c^{\dagger} _{\vec k\sigma}f_{\vec
k^\prime\ga}\tilde s^{*}_{\vec k^\prime -\vec k\Gamma}+H.c.\biggr]$$
\begin{equation} \label{eq:ham}
+\frac{N_s}{2}\sum_{\vec k\Gamma}N_{\Gamma}i\lambda_{\vec
k}\biggl[\sum_{\vec k^\prime}
\tilde s^{*}_{\vec k +\vec k^\prime\Gamma}\tilde s_{\vec
k\Gamma}-q_{\Gamma}\biggr].
\end{equation}
Substituting from equations~\ref{eq:sfluc} and~\ref{eq:lamfluc},
equation~\ref{eq:ham} can be written
in two pieces, one representing the mean field approximation (which we have
solved), and the second
piece coming from the fluctuations in the Bosonic fields, $\delta\tilde s_{\vec
k\Gamma}$ and
$\delta\lambda_{\vec k}$.  The {\em bare} Boson Green functions come from the
terms in
the constraint, the last line of equation~\ref{eq:ham}, which are quadratic in
the fluctuating fields.
Since this involves terms of the form
$\delta\tilde s_{\vec k\Gamma}\delta\tilde s_{\vec k\Gamma}$, $\delta\tilde
s_{\vec k\Gamma}\delta
\lambda_{\vec k}$, and $\delta\lambda_{\vec k}\delta\lambda_{\vec k}$, we write
the bare propagator in a
matrix form (in the static limit),
\begin{equation} \label{eq:barebose}
\hat D^{-1}_{o\Gamma\gp}=-\frac{N_{\Gamma}}{2}\left(\begin{array}{cc}
      i\lambda_o & \tilde s_{o\Gamma} \\ \tilde s_{o\Gamma} & 0
\end{array} \right)\delta_{\Gamma\gp},
\end{equation}
where $\tilde s_{o\Gamma}=s_o/\sqrt{N_{\Gamma}}$.
The specific elements of the matrix are
$D_{oss\Gamma\gp}=-\frac{N_{\Gamma}}{2}i\lambda_o\delta_{\Gamma\gp}$,
$D_{os\lambda\Gamma\gp}=-\frac{N_{\Gamma}}{2}\tilde
s_{o\Gamma}\delta_{\Gamma\gp}$, and
$D_{o\lambda\lambda\Gamma\gp}$=0.

The {\em dressed} Boson Green function will then satisfy a matrix Dyson's
equation,
\begin{equation}  \label{eq:dyson}
\hat D^{-1}_{\Gamma\gp}(\vec q)=\hat D^{-1}_{o\Gamma\gp}(\vec
q)-\hat\Pi_{\Gamma\gp}(\vec q).
\end{equation}
The 2$\times$2 self-energy matrix, $\hat\Pi_{\Gamma\gp}(\vec q)$, due to the
particle-hole excitations,
can be calculated by the usual Feynman diagrammatic techniques.

To motivate the results for $\hat D^{-1}_{\Gamma\gp}$,
consider the diagonal component of the matrix Dyson's equation in Figure 9(a).
Taking all contributions to the
self-energy of order {\em N}$_{\Gamma}$ would give a dressed propagator of
order
1/$N_{\Gamma}$.  In Figure 9(c), the
$\times$ symbols at the corners of the diagrams represent {\em scaled}
hybridization matrix
elements, $\tilde V_{\Gamma\alpha\sigma}(\vec k)$, which are of order one
(${\cal O}(1)$).
Thus, since there is a sum over the degenerate states $\alpha$ of the multiplet
$\Gamma$ and no factor of
$1/N_{\Gamma}$ from the matrix elements to cancel it,  we
say that the Boson self-energy diagram is of order $N_{\Gamma}$.  Technically,
there
is not a free sum over the degeneracy label $\alpha$ of a given multiplet,
because the
hybridization matrix elements, $\tilde V_{\Gamma\alpha\sigma}(\vec k)$, are
dependent upon
$\alpha$.  (That is, we do not have an explicit factor of $N_{\Gamma}$ after
summing over $\alpha$.)
In the $SU(N)$ model, the isotropic hybridization matrix element, $V_o$, is
truly
independent of the degeneracy label, m.  Thus the sum over $m$
yields a factor of N.  In {\em our} case, even though we do not have explicit
factors of
$N_{\Gamma}$, we assume that the closed Fermionic bubbles are the important
diagrams, in analogy
with the $SU(N)$ model.

Figure 9 shows all the unique self-energy diagrams in terms
of the three hybridization-dressed Fermionic Green functions,
equations~\ref{eq:defgf}-\ref{eq:defgmix}.
Evaluation of theses diagrams leads to the following
results for the elements of the (inverse) dressed Bose propagator:
\begin{equation}  \label{eq:predss}
D^{-1}_{ss\Gamma\gp}(\vec q)=2I_{ss\Gamma\gp}(\vec q),
\end{equation}
\begin{equation}  \label{eq:predsl}
D^{-1}_{s\lambda\Gamma\gp}=-i\biggl[\frac{N_{\Gamma}}{2}\tilde
s_{o\Gamma}+\frac{x_{\Gamma}}
{\tilde s_{o\Gamma}}\biggr]\delta_{\Gamma\gp}+iI_{s\lambda\Gamma\gp}(\vec q),
\end{equation}
\begin{equation}  \label{eq:predll}
D^{-1}_{\lambda\lambda\Gamma\gp}(\vec
q)=-\frac{y_{\Gamma}}{T_{o\Gamma}}-\frac{1}{2}I_{\lambda
\lambda\Gamma\gp}(\vec q),
\end{equation}
where the momentum dependent functions are given by
$$I_{ss\Gamma\gp}(\vec q)\equiv\frac{P}{N_s}\sum_{\vec k\vec
k^\prime\alpha\ap}\frac{f(E_{1\vec k})
\tilde s_{o\Gamma}\tilde s_{o\gp}\mu_{\ga\gp\ap}(\vec k)\mu_{\gp\ap\ga}(\vec
k^\prime)
(\epsilon_7-E_{1\vec k})^2(\epsilon_8-E_{1\vec k})^2}{(E_{1\vec
k^\prime}-E_{1\vec k})
(E_{2\vec k^\prime}-E_{1\vec k})(E_{3\vec k^\prime}-E_{1\vec k})(E_{2\vec
k}-E_{1\vec k})}$$
\begin{equation}  \label{eq:preiss}
\times\frac{\bigl[\delta_{\vec k^\prime,\kq}+\delta_{\vec k^\prime,\vec k -\vec
q}\bigr]}
{(\epsilon_{\Gamma}-E_{1\vec k})(\epsilon_{\gp}-E_{1\vec k})(E_{3\vec
k}-E_{1\vec k})},
\end{equation}
$$I_{s\lambda\Gamma\gp}(\vec q)\equiv\frac{P}{N_s}\sum_{\vec k\vec
k^\prime}\frac{f(E_{1\vec k})
(\epsilon_7-E_{1\vec k})^2(\epsilon_8-E_{1\vec k})^2}{(E_{2\vec k}-E_{1\vec k})
(E_{3\vec k}-E_{1\vec k})(E_{1\vec k^\prime}-E_{1\vec k})(E_{2\vec
k^\prime}-E_{1\vec k})
(E_{3\vec k^\prime}-E_{1\vec k})}$$
$$\times\sum_{\alpha\ap}\frac{\tilde s_{o\Gamma}\tilde s_{o\gp}}{(E_{1\vec
k}-\epsilon_
{\Gamma})(E_{1\vec k}-\epsilon_{\gp})}\biggl[\frac{\tilde
s_{o\Gamma}\mu_{\ga\gp\ap}(\vec k)
\mu_{\gp\ap\ga}(\vec k^\prime)}{(E_{1\vec k}-\epsilon_{\Gamma})}$$
\begin{equation}  \label{eq:preisl}
+\frac{\tilde s_{o\gp}\mu_{\gp\ap\ga}(\vec k)\mu_{\ga\gp\ap}(\vec
k^\prime)}{(E_{1\vec k}
-\epsilon_{\gp})}\biggr]\bigl[\delta_{\vec k^\prime,\kq}+\delta_{\vec
k^\prime,\vec k -\vec q}\bigr],
\end{equation}
$$I_{\lambda\lambda\Gamma\gp}(\vec q)\equiv\frac{P}{N_s}\sum_{\vec k\vec
k^\prime\alpha\ap}\frac{f(E_{1\vec k})
\tilde s^{2}_{o\Gamma}\tilde s^{2}_{o\gp}\mu_{\ga\gp\ap}(\vec
k^\prime)\mu_{\gp\ap\ga}(\vec k)
\bigl[\delta_{\vec k^\prime,\vec k -\vec q}+\delta_{\vec
k^\prime,\kq}\bigr]}{(E_{1\vec k^\prime}
-E_{1\vec k})(E_{2\vec k^\prime}-E_{1\vec k})(E_{3\vec k^\prime}-E_{1\vec
k})(E_{2\vec k}-E_{1\vec k})}$$
\begin{equation}  \label{eq:preill}
\times\frac{(\epsilon_7-E_{1\vec k})^2(\epsilon_8-E_{1\vec
k})^2}{(\epsilon_{\Gamma}-E_{1\vec k})^2
(\epsilon_{\gp}-E_{1\vec k})^2(E_{3\vec k}-E_{1\vec k})}.
\end{equation}
In equations~\ref{eq:preiss} -~\ref{eq:preill}, $E_{n\vec k}$ are the
quasiparticle band energies.
The quasiparticle chemical potential was chosen to lie in the lowest band, such
that $f(E_{n\vec k})$
is nonzero only for $n=1$.  (Recall this is zero temperature calculation.)
The anisotropic function $\mu_{\ga\gp\ap}(\vec k)$ is defined in
equation~\ref{eq:mu}.
The parameters $x_{\Gamma}$ and $y_{\Gamma}$ are numbers which depend on the
mean field parameter set
used.  See Table 3 for the values of these parameters.

Note the presence of the principal value integrals over the Brillouin zone.  We
are forced to
evaluate the integrals  numerically, as discussed in the next section.  This is
the most
labor-intensive part of the calculation.

\begin{center}
{\Large VIII. ANALYTIC TETRAHEDRON METHOD}
\end{center}
\vspace{0.4cm}

We wish to evaluate the three dimensional principal value integrals that have
arisen in the calculation of the Bosonic self-energy, {\em i.e.} the functions
$I_{ss\Gamma\gp}(\vec q)$, $I_{s\lambda\Gamma\gp}(\vec q)$, and
$I_{\lambda\lambda\Gamma\gp}(\vec q)$ of
equations~\ref{eq:preiss}-\ref{eq:preill}.  To do so, we use a
procedure developed originally to calculate the real part of spectral functions
of the form\cite{Jepsen 71},\cite{Lehmann 70},\cite{Rath 75},\cite{Lingard 75},
\begin{equation}  \label{eq:respec}
\Phi^\prime(E)=Re\Phi(E)=P\frac{1}{N_s}\sum_{\vec k}\frac{M(\vec k)}{E-E_{n\vec
k}}.
\end{equation}
$\Phi^\prime$ looks like the real part of a dynamic susceptibility.  Handled
numerically, the
principal value nature of the sum in equation~\ref{eq:respec} makes it
non-trivial.  In fact, if we
tried to do the sum by simply evaluating the function at many points throughout
the zone, and then
multiplying by a weighting factor, we would find an essentially infinite
variance.  That is, upon
averaging the result over many different mesh sizes, the variance of the mean
would be huge compared
to the mean-value itself\cite{Jacobs 71}.

The analytic tetrahedron method\cite{Rath 75},\cite{Lingard 75} evaluates
expressions
like that of equation~\ref{eq:respec} by breaking up the Brillouin zone into
tetrahedra, where the band energies $E_{n\vec k}$ need to be known only at the
four corners of a given tetrahedron.  Any band energy for $\vec k$ inside a
tetrahedron is
interpolated from the energies at the corners.  Within this assumption of
linear
interpolation, and also assuming the effective matrix element, $M(\vec k)$, is
constant
inside the tetrahedron, Rath and Freeman\cite{Rath 75}, and independently
Ling\aa rd
\cite{Lingard 75} showed that the principal value integral over the tetrahedron
could be
performed analytically.

We write our functions $I_{ab\Gamma\gp}(\vec q)$ in the following form:
\begin{equation}  \label{eq:specsigma}
I_{ab\Gamma\gp}(\vec q)=\frac{P}{N_s}\sum_{\vec k\vec k^\prime}
\frac{f(E_{1\vec k})M_{ab\Gamma\gp}(\vec k,\vec k^\prime)}
{E_{1\vec k^\prime}-E_{1\vec k}}\delta_{\kp,\vec k\pm\vec q},
\end{equation}
where the matrix elements are a complicated function of $\vec k$ and
$\vec q$ which come from equations~\ref{eq:preiss},~\ref{eq:preisl},
and~\ref{eq:preill}.
That is, direct comparison of equation~\ref{eq:specsigma} with
equations~\ref{eq:preiss}-\ref{eq:preill}
gives the structure of the matrix elements $M_{ab\Gamma\gp}\vec k, \vec
k^\prime$.
Note that {\em ab}
can represent {\em ss}, $s\lambda$, or $\lambda\lambda$.
The Fermi function,
of course, restricts us to only the volume below the Fermi surface.  Since the
surfaces of constant
energy are planar, we need to know the possible unique ways a plane can cut
through a tetrahedron.
It turns out, fortunately, that there are only three such ways; and
furthermore, in each of the three
cases, the two subdivided volumes of the intersected tetrahedron (the volume
above the Fermi surface
and the volume below the Fermi surface) are themselves either a single
tetrahedron or a composite
of three tetrahedra.  We {\em never} deal with anything but tetrahedra.  So it
is just a matter of
geometry to find out which tetrahedra are beneath the Fermi surface and hence
contribute to the
integral of equation~\ref{eq:specsigma}.

To evaluate our self-energies, we need the analytic result for the following
integral over a tetrahedron:
\begin{equation}  \label{eq:tetint}
I=\int_{tetra}\frac{d^3k}{E_{1\vec k^\prime}-E_{1\vec k}},
\end{equation}
where $\vec k^\prime=\vec k \pm \vec q$.
For simplicity of notation we define the energy differences
\begin{equation}  \label{eq:Vi}
V_i\equiv E_{1\vec k^\prime i}-E_{1\vec k i},
\end{equation}
where the ``1'' signifies the band index, and $i=1,2,3,4$ labels the four
corners of the
tetrahedron.
The result for the integral $I$ depends only on the $V_i$ and the volume of the
tetrahedron.
Using the
notation of Rath and Freeman\cite{Rath 75} the results is:
\begin{equation}  \label{eq:iresult}
I=3v_{tet}\biggl(\frac{V^{2}_{1}}{D_1}\ln\bigg|\frac{V_1}{V_4}\bigg|
+\frac{V^{2}_{2}}{D_2}\ln\bigg|\frac{V_2}{V_4}\bigg|
+\frac{V^{2}_{3}}{D_3}\ln\bigg|\frac{V_3}{V_4}\bigg|\biggr),
\end{equation}
where $v_{tet}$ is the volume of the tetrahedron, and where
\begin{eqnarray}
  D_1= & (V_1-V_4)(V_1-V_3)(V_1-V_2) \label{eq:d1}\\
  D_2= & (V_2-V_4)(V_2-V_3)(V_2-V_1) \label{eq:d2}\\
  D_3= & (V_3-V_4)(V_3-V_2)(V_3-V_1). \label{eq:d3}
\end{eqnarray}
We have studied equation~\ref{eq:iresult} carefully in the limits where some of
the $V_i$s are equal
to each other or are equal to zero.  It turns out that there are quite a few
such cases,  but for the
sake of space, we do not tabulate them here\cite{Trees thesis}.

Note that these three dimensional integrals can have an entire surface of poles
inside
the Brillouin zone that must be handled properly.
In contrast, a one-dimensional principal value integral can be regularized
numerically by
basically subtracting off the divergence\cite{Cox 85}, and such a procedure is
facilitated by the
relatively small number of poles throughout the domain of integration.  In our
three-dimensional
integral, if $n^{3}_{mesh}$ is the number of sub-cubes inside the cubic
Brillouin zone, then
there are of the order of $n_{mesh}^2$ poles, which clearly gets large as
$n_{mesh}$ increases,
and it is no longer possible
to regularize the integral in a simple way.  The large quantity of work done by
(mostly
electronic structure) physicists in this field of numerical k-space sums, is
indicative of the
degree of complexity inherent to these problems \cite{Gilat 72},\cite{Gilat}.

We now show the results for the self-energy functions
$I_{ss}(\vec q)$, $I_{s\lambda}(\vec q)$, and $I_{\lambda\lambda}(\vec q)$,
which have been summed
over the crystal field multiplet indices,
\begin{equation}  \label{eq:newestiss}
I_{ss}(\vec q)\equiv \sum_{\Gamma\gp}\frac{I_{ss\Gamma\gp}(\vec
q)}{\sqrt{N_{\Gamma}}
\sqrt{N_{\gp}}},
\end{equation}
\begin{equation}  \label{eq:newestisl}
I_{s\lambda}(\vec q)=\sum_{\Gamma\gp}\biggl(\frac{I_{s\lambda\Gamma\gp}(\vec
q)}
{\sqrt{N_{\Gamma}}}
+\frac{I_{s\lambda\Gamma\gp}(\vec q)}{\sqrt{N_{\gp}}}\biggr),
\end{equation}
and
\begin{equation}  \label{eq:newestill}
I_{\lambda\lambda}(\vec q)=\sum_{\Gamma\gp}I_{\lambda\lambda\Gamma\gp}(\vec q).
\end{equation}
For a given mesh parameter, $n_{mesh}$,
the total number of tetrahedra in the Brillouin zone is 8$\times n^{3}_{mesh}$.
The functions $I_{ss}$, $I_{s\lambda}$, and $I_{\lambda\lambda}$ are plotted as
a function of the
mesh parameter, $n_{mesh}$, in Figures 10-12 for mean field parameter set (a).
The results for parameter set (b) are similar.
We chose the momentum $q=0.5\hat z$, measured in units of $\pi/a$, where $a$ is
the lattice spacing.
(We take $a=$3.89\AA.)  The matrix elements are assumed constant inside
a given tetrahedron.  Since they are complicated anisotropic functions,
however, the matrix elements
can vary a great deal throughout the Brillouin zone,
and thus they cause fluctuations in the value of the $I_{ab}(\vec q)$
as a function of $n_{mesh}$.  This is clearly visible in figures 10-12.

It is clear that if we need convergence of the Green functions
to several decimal places, it would require a mesh with $n_{mesh}>80$.
In background, a calculation for $n_{mesh}=40$ takes about
one hour on a DEC 5100. Since the computing time goes as
$n_{mesh}^3$, it is clear that a run with $n_{mesh}=80$ requires a long run
time.
Attempts at vectorization of the code were hindered by a plethora of logic
statements required for the subroutine that performed the integral in
equation~\ref{eq:tetint}.
Actually, this numerical integration is parallel in nature.  Each of the
$n_{mesh}^3$
subcubes could, in principle, be integrated independently of the others,
and at the end the net result would be the
sum of the results from each subcube.  Naively, this is the kind of problem a
parallel machine should be able
to handle well.

It is also interesting to note that calculations of the susceptibility or
dielectric constant based
on electronic structure data, generally are not performed for $n_{mesh}$
greater than about 30
\cite{Matthew}.  The complicated nature of our matrix elements have forced us
to push the procedure
to very large values (by anyone's standards) of $n_{mesh}$.  We shall discuss
in the next section,
how our conclusions on pairing instabilities in the infinite-U Anderson lattice
take into account
this slow convergence of the Bosonic self-energy.

\begin{center}
{\Large IX. QUASIPARTICLE SCATTERING AMPLITUDE}
\end{center}
\vspace{0.4cm}

In this section, we present our analysis of the scattering amplitude
$\Gamma_{QP}(\vec k,\kp)$,
which allows for quasiparticle interactions via the exchange of 4f density
fluctuations.
The exchanged Boson will be represented by the (dressed to order 1/N) Boson
matrix Green function
calculated in Sec. VIII.  The diagrams for the scattering amplitude are
presented in Figure 13, where
the straight lines represent the incoming (and outgoing) quasiparticles and the
wavy line is an element of
the (matrix) Boson Green function, $\hat D_{\Gamma\gp}(\vec q)$.  The large
black circles represent
quasiparticle vertices, $\gamma_{f\Gamma}(\vec k, \kp)$ and
$\gamma_{mix\Gamma}(\vec k, \kp)$, which
are calculated by writing the full Hamiltonian in the quasiparticle basis:
$$H_{QP}=\sum_{\vec k\sigma nn^\prime}\xi_{\vec k}A^{*}_{n}(\vec
k)A_{n^\prime}(\vec k)
Q^{\dagger}_{\vec k n\sigma}Q_{\vec k n^\prime\sigma}$$
$$+\sum_{\vec k\vec k^\prime \ga
nn^\prime\sigma\sigma^\prime}\gamma_{f\Gamma}(\vec k, \vec k^\prime)
Q^{\dagger}_{\vec k n\sigma}\biggl[E_{\Gamma}\delta_{\vec k\vec
k^\prime}+i\lambda_{\vec k -
\vec k^\prime}\biggr]Q_{\vec k^\prime n^\prime\sigma^\prime}$$
\begin{equation}  \label{eq:newhqp}
+\sum_{\vec k\vec k^\prime\ga\sigma
nn^\prime\sigma^\prime}\biggl[\gamma_{mix\Gamma}(\vec k, \vec k^\prime)
Q^{\dagger}_{\vec k n\sigma}Q_{\vec k^\prime n^\prime \sigma^\prime}\tilde
s^{*}_{\vec k^\prime -
\vec k \Gamma} + H.c.\biggr]+H_{constraint}.
\end{equation}
Recall that $Q_{\vec k n\sigma}$ destroys a quasiparticle of momentum $\vec k$,
pseudospin $\sigma$, and band index $n$.
The vertex functions $\gamma_f$ and $\gamma_{mix}$ thus come from the unitary
transformation that
diagonalizes $H_{MF}$ (see equations~\ref{eq:candq} and~\ref{eq:fandq})
\begin{equation} \label{eq:fvertex}
\gamma_{f\Gamma}(\vec k,\vec k^\prime)\equiv \frac{A^{*}_{n}(\vec
k)A_{n^\prime}(\vec k^\prime)
\tilde s^{2}_{o\Gamma}\tilde V^{*}_{\ga\sigma^\prime}(\vec k^\prime)\tilde
V_{\sigma\ga}
(\vec k)}{\bigl(\epsilon_{\Gamma}-E_{n\vec k}\bigr)
\bigl(\epsilon_{\Gamma}-E_{n^\prime\vec k^\prime}\bigr)},
\end{equation}
\begin{equation} \label{eq:mixvertex}
\gamma_{mix\Gamma}(\vec k,\vec k^\prime)\equiv -\frac{A^{*}_{n}(\vec
k)A_{n^\prime}(\vec k^\prime)
\tilde s_{o\Gamma}\tilde V^{*}_{\ga\sigma^\prime}(\vec k^\prime)\tilde
V_{\ga\sigma}(\vec k)}
{\epsilon_{\Gamma}-E_{n^\prime\vec k^\prime}}.
\end{equation}

As shown below, we project $\Gamma_{QP}(\vec k, \kp)$ onto states of cubic
symmetry, $\Phi_{\eta}(
\vec k)$, the so-called cubic harmonics, where $\eta$ labels the irreducible
representations of the
octahedral group ${\cal O}_h$.
The product is then averaged over the Fermi surface,
\begin{equation}  \label{eq:defineaverage}
\Gamma_{\eta}=\int\frac{d\hat k}{4\pi}\int\frac{d\hat k^\prime}{4\pi}
\Phi^{*}_{\eta}(\hat k^\prime)\Gamma_{QP}(\vec k, \vec
k^\prime)\Phi_{\eta}(\hat k).
\end{equation}
A superconducting instability of symmetry $\eta$ is signaled by a negative
value
of the corresponding average, $\Gamma_{\eta}$.
See Table 4 for a list of the cubic harmonics used in this calculation.
The character table for the octahedral group ${\cal O}$ is presented in Table
5, where the irreducible
representations are listed:  $A_1$, $A_2$, $E$, $T_1$, and $T_2$.  The group
${\cal O}_h$ follows
from the group ${\cal O}$ by including inversions.
This means the representations pick up a
subscript $''g''$ or $''u''$ depending on if
the they are even or odd, respectively, under parity.  Because of the
complexity
in calculating the dressed bose propagators we have assumed a spherical Fermi
surface for the average in equation~\ref{eq:defineaverage}.  We do not feel
this is
a weakness of the calculation for reasons discussed elsewhere\cite{Trees 94}.

In this paper, we shall discuss only even parity
pairing states.  This restriction is based on the following experimental
evidence for \cc:
the need for strong Pauli limiting to fit the low temperature upper critical
field data,
$H_{c2}(0)$\cite{Steglich 85}; the reduced spin susceptibility below $T_c$ as
measured by the
$^{63}$Cu Knight shift\cite{Asayama 88}, and the observed $T^{3}$ temperature
dependence of
the nuclear-spin relaxation rate below $T_c$\cite{Asayama 88}.  The strong
Pauli limiting actually
only argues against equal spin pairing states.  As Ueda and Rice
showed\cite{Ueda 85}, in the
presence of spin-orbit coupling, Pauli limiting is possible for pairing states
of $T_{1u}$ or
$T_{2u}$ symmetry.  (Both of these odd-parity states would have a gap with
point nodes as opposed to
line nodes on the Fermi surface.)
These experimental facts put together, however, might be considered
reasonable evidence for even-parity pairing in \cc.


We have found it useful to study the properties of the scattering amplitude in
two steps.  First,
by setting the functions $I_{ss\Gamma\gp}(\vec q)$, $I_{s\lambda\Gamma\gp}(\vec
q)$, and
$I_{\lambda\lambda\Gamma\gp}(\vec q)$ to zero, we simplify the problem
considerably to that of
two quasiparticles scattering via exchange of a momentum independent Boson.  In
real space, this
corresponds to a local interaction between the quasiparticles.
We find these local interactions, when averaged over the Fermi surface, are
substantially different
from those calculated within the jellium model by Zhang and T. K.
Lee\cite{Zhang 87}.

In the case of cubic symmetry, inclusion of the functions $I_{ss\Gamma\gp}(\vec
q)$,
$I_{s\lambda\Gamma\gp}
(\vec q)$, and $I_{\lambda\lambda\Gamma\gp}(\vec q)$ is the computationally
intensive part of
this calculation.  If there are strong local {\em repulsive} interactions in
the $\eta$
pairing channel, then the only way
to get a pairing instability ({\em i.e.} $\Gamma_{\eta}<0$), is to have the
functions
$I_{ss\Gamma\gp}(\vec q)$, $I_{s\lambda\Gamma\gp}(\vec q)$, and
$I_{\lambda\lambda\Gamma\gp}(\vec q)$,
which represent the effect of non-local
interactions, overcome the repulsion.  Zhang and Lee discovered that in
spherical symmetry
these q-dependent contributions are too weak to overcome the local repulsions
in the $s$, $d$, and
$g$-wave pairing states.  We find that
in cubic symmetry (with crystal-field splitting)
attractive nonlocal interactions can overpower local repulsions in the
$T_{1g}$ pairing channel, thus giving evidence for a $T_{1g}$ pairing
instability.

When we evaluate the diagrams of Figure 13, we can write the quasiparticle
scattering amplitude in the even-parity (pseudospin singlet) channel as
$$\Gamma_{QP}(\vec k,\kp)=
\frac{1}{4}\sum_{\ga\gp\ap\sigma\sigma^\prime}\frac{A^{2}_{1}(\vec
k)A^{2}_{1}(\kp)
\tilde s_{o\Gamma}\tilde s_{o\gp}\tilde V^{*}_{\ga\sigma^\prime}(\kp)\tilde
V_{\sigma^\prime
\gp\ap}(\kp)\tilde V^{*}_{\gp\ap\sigma}(\vec k)\tilde V_{\sigma\ga}(\vec
k)}{(\epsilon_{\Gamma}
-E_{1\vec k})(\epsilon_{\gp}-E_{1\vec k})}$$
$$\times\biggl[D_{ss\Gamma\gp}(\kp -\vec k)+D_{ss\Gamma\gp}(-\kp -\vec
k)-i\frac{\tilde s_{o\Gamma}}
{\epsilon_{\Gamma}-E_{1\kp}}\biggl(D_{s\lambda\gp}(\kp -\vec
k)+D_{s\lambda\gp}(-\kp -\vec k)
\biggr)$$
$$-i\frac{\tilde
s_{o\gp}}{\epsilon_{\gp}-E_{1\kp}}\biggl(D_{s\lambda\Gamma}(\kp -\vec k)
+D_{s\lambda\Gamma}(-\kp -\vec k^\prime)\biggr)$$
\begin{equation}  \label{eq:finalg}
-\frac{\tilde s_{o\Gamma}\tilde
s_{o\gp}}{(\epsilon_{\Gamma}-E_{1\kp})(\epsilon_{\gp}-E_{1\kp})}
\biggl(D_{\lambda\lambda}(\kp-\vec k)+D_{\lambda\lambda}(-\kp -\vec
k)\biggr)\biggr],
\end{equation}
where the symbols mean the following:

\noindent
$\bullet$$\tilde V_{\sigma\ga}(\vec k)=\sqrt{N_{\Gamma}}V_{\sigma\ga}$ is the
hybridization matrix
element between a plane wave
conduction state of momentum $\vec k$ and spin $\sigma$ and a crystal field
state with quantum
numbers $\Gamma$ and $\alpha$.
$N_{\Gamma}$ labels the degeneracy of the $\Gamma$ multiplet.

\noindent
$\bullet$$\tilde s_{o\Gamma}=s_o/\sqrt{N_{\Gamma}}$ can be thought of as the
mean field
hybridization renormalization coefficient.
At mean field level, the bare hybridization is renormalized (due to the
constraint of only
allowing hopping onto an empty $4f$ site) to the value $\tilde
s_{o\Gamma}\tilde V_{\sigma\ga}$.

\noindent
$\bullet$$\epsilon_{\Gamma}$ is the self-consistent, shifted mean field energy
of the $\Gamma$ multiplet.

\noindent
$\bullet$$E_{1\vec k}$ is the quasiparticle energy for the lowest, or first,
band as a function of momentum.

\noindent
$\bullet$$A^{2}_{1}(\vec k)$ is the quasiparticle normalization function
defined in equation~\ref{eq:fulla},
and which is {\em very} strongly peaked at the six points where the Brillouin
zone axes intersect
the Fermi surface.

\noindent
$\bullet$The components of the Bosonic Green function are:
$$D_{ss\Gamma\gp}(\vec q, \tau)\equiv\langle \delta\tilde s_{-\vec
q\Gamma}(\tau)\delta\tilde s_{\vec q\gp}(0)\rangle;$$
$$D_{s\lambda\Gamma}(\vec q, \tau)\equiv\langle \delta\tilde s_{-\vec
q\Gamma}(\tau)\delta\lambda_{\vec q}(0)\rangle;$$
$$D_{\lambda\lambda}(\vec q, \tau)\equiv\langle\delta\lambda_{-\vec
q}(\tau)\delta\lambda_{\vec q}(0)\rangle.$$
Note that we are taking the static limit of these Green functions.
It became clear
to us that including the frequency dependence of the boson Green functions was
impossible,
given the difficult numerical integrals encountered even in the static limit.

When evaluated on the Fermi surface (at zero temperature),
the band energies $E_{1\vec k}$ and $E_{1\kp}$ are set equal to the
quasiparticle
chemical potential, $\mu$.  Thus a term like $\epsilon_{\Gamma}-E_{1\vec k}$
becomes
\begin{equation}  \label{eq:defKtemp}
\epsilon_{\Gamma}-\mu=T_{o\Gamma},
\end{equation}
which is the Kondo temperature of the $\Gamma$ multiplet.
For the case of \cc, the
Kondo temperature of the $\Gamma_7$ doublet is approximately 10 K,
while for the $\Gamma_8$ quartet,
\begin{equation}  \label{eq:To8again}
T_{o8}=T_{o7}+\Delta_{CEF}=370 K.
\end{equation}
This explains how to treat all the energy denominators
in equation~\ref{eq:finalg}.

As mentioned, it is instructive to consider the so-called local limit of
equation~\ref{eq:finalg} in which the
self-energy functions $I_{ss}(\vec q)$, $I_{s\lambda}(\vec q)$, and
$I_{\lambda\lambda}(\vec q)$ are
set to zero,
$$\Gamma_{local}(\vec
k,\kp)=\frac{1}{4}\sum_{\ga\gp\ap\sigma\sigma^\prime}\frac{A^{2}_{1}(\vec k)
A^{2}_{1}(\kp)\tilde s_{o\Gamma}\tilde s_{o\gp}\tilde
V^{*}_{\ga\sigma}(\kp)\tilde V_{\sigma^\prime
\gp\ap}(\kp)\tilde V^{*}_{\gp\ap\sigma}(\vec k)\tilde V_{\sigma\ga}(\vec
k)}{T_{o\Gamma}T_{o\gp}}$$
\begin{equation}  \label{eq:glocal}
\times 2\biggl[\frac{\tilde
s_{o\Gamma}}{\sqrt{N_{\gp}}T_{o\Gamma}\Gamma_{os\lambda}}+
\frac{\tilde s_{o\gp}}{\sqrt{N_{\Gamma}}T_{o\gp}\Gamma_{os\lambda}}-
\frac{\Gamma_{o\lambda\lambda}}{\sqrt{N_{\Gamma}N_{\gp}}\Gamma^{2}_{os\lambda}}\biggr],
\end{equation}
where
\begin{equation}  \label{eq:gammasl}
\Gamma_{os\lambda}=\sum_{\Gamma}\biggl(\frac{1}{2}N_{\Gamma}\tilde
s_{o\Gamma}+\frac{x_{\Gamma}}
{\tilde s_{o\Gamma}}\biggr),
\end{equation}
\begin{equation}  \label{eq:gammall}
\Gamma_{o\lambda\lambda}=\sum_{\Gamma}\frac{y_{\Gamma}}{T_{o\Gamma}}.
\end{equation}
The combination of normalization functions, $A^{2}_{1}(\vec k)A^{2}_{1}(\kp)$,
and the product of the four hybridization matrix elements is due to the
anisotropic vertices $\gamma_{mix\Gamma}(\vec k,\kp)$ and $\gamma_{f\gp}(\vec
k,\kp)$.  The
contribution from the Bose Green function is that which remains inside the
square brackets
in equation~\ref{eq:glocal}.

The local scattering amplitude of equation~\ref{eq:glocal} is much simpler to
deal with than the
full expression of equation~\ref{eq:finalg}.
Roughly, the functions $I_{ss}(\vec q)$, $I_{s\lambda}(\vec q)$, and
$I_{\lambda\lambda}(\vec q)$
can be thought of as renormalizing the local
interactions.
In the next section, we present our results for the Fermi surface averaged
scattering
amplitude, $\Gamma_{\eta}$, both with and without the Bosonic self energy.

\begin{center}
{\Large X. RESULTS-LOCAL LIMIT}
\end{center}
\vspace{0.4cm}

We now present our results for the Fermi surface averaged (local) quasiparticle
interactions,
$\Gamma_{local,\eta}$, where $\eta$=$A_{1g}$, $E_g$, $T_{1g}$, or $T_{2g}$.
We have not included pairing states in the $A_{2g}$ representation (See Tables
4 and 5) for two reasons:
(1) the lowest order
spherical harmonic present in the cubic harmonic $\Phi_{A_{2g}}$ is
$Y_{6m}$\cite{Altmann 65},
and the relatively rapid variation throughout the
Brillouin zone of the pair wavefunction would correspond to a pairing state
with
a high kinetic energy and hence should be
less accessible than the pairing states labeled by the other representations;
(2) the $A_{2g}$ state would also
require a finer mesh for the Fermi surface average than the one we have used
and hence would further increase the
(already considerable) overall computing time.

Table 6 gives the ratio $\Gamma_{local,\eta}/T_{o7}$ for both mean-field
parameter sets (a) and (b). Both parameter sets have Kondo temperatures of
about 10 K.
Also presented in table 6 is the contribution to the Fermi surface average from
the so-called
``hot-spots''.  For the hot-spot contribution, the normalization functions
$A_{1}^{2}(\vec k)$
and $A_{1}^{2}(\vec k^{\prime})$ in
equation~\ref{eq:glocal} were
approximated with delta functions
in k-space that sampled only the six points on the Fermi surface intersected by
the
Brillouin zone axes.  The integrated weight of the delta functions was chosen
to equal the area under
the peaks, as shown in figure 4.  Such a calculation gives us a feeling for the
importance of
these special points where the normalization function, $A_{1}(\vec k)$, is
rapidly changing.
This contribution is labeled by $\Gamma_{local,hot}$ in Table 6.
We see from Table 6 that

\noindent
$\bullet$ The $A_{1g}$ pairing channel
has (by-far) the largest (repulsive) local interaction, but it is not dominated
by what
happens at the ``hot-spots''.

\noindent
$\bullet$ In the $E_g$ channel, the strong anisotropy
of the normalization function, gives rise to a weakly {\em attractive} local
interaction.
In fact, the attractive local interaction here is dominated by the
contributions from the
hot-spots.  We note from Table 5 that a pairing state of $E_g$ symmetry
transforms like
$x^2-y^2$ or $3z^2-r^2$, which have maxima along the
directions of the axes.
The normalization is also
{\em strongly} peaked along the axes.  Thus the $E_g$ states are greatly
affected by $A^{2}_{1}(\vec k)$.

\noindent
$\bullet$ In the $T_{1g}$ and $T_{2g}$ channels, the local interaction is weak
and repulsive.
Neither symmetry channels can ``see'' what happens at the ``hot-spots'',
because both
cubic harmonics, $\Phi_{T1g}$ and $\Phi_{T2g}$ vanish identically at these
points.

In Table 7 we present for comparison, the results for the Fermi-surface
averaged scattering amplitude
from Zhang and Lee's\cite{Zhang 87} jellium model
calculation, where Legendre polynomials ($P_{\eta}$) play the role of the cubic
harmonics.
States described by $\eta$=0 are s-wave;
$\eta$=2 corresponds to d-wave; and $\eta$=4 is g-wave.  Zhang and Lee find all
non-zero
interactions are repulsive and of about the same strength. As we have said,
they also found
the inclusion of non-local interactions
was not sufficient to overcome the local repulsions.
In octahedral symmetry, however, the very weak
local interactions in the $E_g$, $T_{1g}$, and $T_{2g}$ pairing channels make a
superconducing
instability likely.

\begin{center}
{\Large XI. RESULTS-INCLUDING NON-LOCAL INTERACTIONS}
\end{center}
\vspace{0.4cm}

This section discusses our results for the Fermi surface average of the
full quasiparticle scattering amplitude, $\Gamma_{QP}(\vec k, \kp)$, as given
in equation~\ref{eq:finalg}.  The functions
$I_{ss\Gamma\gp}(\vec q)$, $I_{s\lambda\Gamma\gp}(\vec q)$, and $I_{\lambda
\lambda\Gamma\gp}(\vec q)$ (see equations~\ref{eq:preiss}-\ref{eq:preill}),
contain the physics of
the screened (by density fluctuations
of the coupled conduction-$4f$ electrons) slave Bosons.  The momentum
dependence represents the contribution of non-local quasiparticles
interactions in real space.

{}From equations~\ref{eq:finalg} and \ref{eq:defineaverage},
we see that we must evaluate the self-energy functions
for all unique combinations of $\vec k\pm\kp$.
This we do separately from the actual averaging process.  We store the
required values for $I_{ss\Gamma\gp}(\vec q)$, $I_{s\lambda\Gamma\gp}(\vec q)$,
and
$I_{\lambda\lambda\Gamma\gp}(\vec q)$ in a
look-up table.

We discussed the general convergence properties of the functions
$I_{ss\Gamma\gp}(\vec q)$,
$I_{s\lambda\Gamma\gp}(\vec q)$
, and $I_{\lambda\lambda\Gamma\gp}(\vec q)$ in Sec. VIII, where we saw that,
due to the complicated
effective matrix elements in cubic symmetry (see equation~\ref{eq:specsigma}),
they are very slow to converge as a function of the Brillouin zone mesh size.
The slow convergence prompted us to try the following line of attack.
We have averaged the scattering
amplitude (equations~\ref{eq:finalg}) over the Fermi surface with a fixed
{\em averaging} mesh.
We vary, however, the mesh for calculating the Bosonic Green function, as
characterized by the parameter $n_{mesh}$, and study the {\em average}
interactions as a function of $n_{mesh}$.
With this procedure, it became clear that $n_{mesh}\approx$ 50 is a practical
limit of the mesh size.  Creating the look-up table for the Bosonic
self-energies would require well over a week of runtime on a DEC5100 for
anything bigger.  Thus, it is important to ask if we can make any conclusions
about possible pairing instabilities for $n_{mesh}\leq$ 50.

The values of $\Gamma_{\eta}/T_{o7}$  for $\eta=A_{1g}$
and $\eta=T_{2g}$ are plotted as a function of $n_{mesh}$ in
Figure 14 for parameter set (a) and in Figure 16
for parameter set (b).  Also on the plots, the local value for the average are
marked
by horizontal lines for each representation (except $A_{1g}$).
The results show the following:

\noindent
$\bullet$ In the $A_{1g}$ channel, the interactions are clearly repulsive.
 The non-local interactions, however, are attractive, since the average
 {\em local} repulsion is of the size $\Gamma_{local,A1g}/T_{o7}$=3.54,
and the inclusion of the momentum dependent Bosonic self-energy gives
$\Gamma_{A1g}/T_{o7}\approx$ 1.2.  Thus the non-local
contribution has reduced the local repulsion by about a factor of two.

\noindent
$\bullet$ In the $T_{2g}$ channel, but for a glitch at $n_{mesh}$=23,
the interactions are repulsive for parameter set (a), with $\Gamma_{T2g}/T_{o7}
\approx$0.20.  For parameter set (b) (Figure 16), however, it is less clear if
the average interaction is attractive or repulsive.  Taking the results for
both
parameter sets together, we believe there is {\em no} pairing instability of
$T_{2g}$ symmetry.

The values of $\Gamma_{\eta}/T_{o7}$ for $\eta=E_g$ and
$\eta=T_{1g}$ are plotted in Figure 15 for parameter set (a) and in
Figure 17 for parameter set (b).

\noindent
$\bullet$The results for $T_{1g}$ show attractive interactions, with an average
value $\Gamma
_{T1g}/T_{o7}$=-0.212$\pm$0.025 for parameter set (a), and =-0.226$\pm$0.049
for parameter
set (b).

\noindent
$\bullet$  The interactions in the $E_g$ channel are less well behaved, but
are attractive for 25$\leq n_{mesh}\leq$ 47 (for parameter set (a)).
These numbers point to a
possible superconducting instability of $E_g$ symmetry, but the results have
not converged enough for us to be sure.  As a rough guide, using the numbers
for 25$\leq n_{mesh}\leq$ 47 would give $\langle\Gamma\rangle_{Eg}/T_{o7}
\approx$-0.163$\pm$0.054.  The values for parameter set (b), however,  have a
surprisingly
large fluctuation at $n_{mesh}$=31, which makes it very difficult to make any
conclusion
for $n_{mesh}\leq$ 35.  It is clear that the $E_g$ states are the most
sensitive to the fluctuations
due to the anisotropic matrix elements in the Bosonic self-energies.
{}From these results, we see that the $T_{1g}$ pairing state
is the most likely candidate for a pairing instability in cubic symmetry.

\begin{center}
{\Large XII. DISCUSSION}
\end{center}
\vspace{0.4cm}

As we have said, $n_{mesh}\approx$50 is a practical limit on the size of the
Bosonic self-energy mesh that we could run on a local workstation.
The appearance of fluctuations in the averaged scattering amplitude (Figures
14-17),
which limit our conclusions about possible pairing instabilities in cubic
symmetry, is not shocking.
After all, we know that there are fluctuations in the Bosonic self-energies
for $n_{mesh}\approx$80 (See figures 11-13).  As we discussed, such variations
are
due to the anisotropic matrix elements in the principal value integrals over
the Brillouin
zone.  The zone is divided into a large number of tetrahedra
(8$\times(n_{mesh})^3$ to be exact),
and the matrix elements are assumed constant inside a given tetrahedron.  If
the matrix elements
are sharply peaked in some region of the Brillouin zone, then it is easy to see
how
fluctuations can occur.  If the mesh is constructed so that the matrix elements
are evaluated
very near a peak, then only a small shift in the mesh is required before the
matrix elements
will be evaluated at a point far down on the sides of the peak.  Thus a small
change in mesh
size could result in a large change in the evaluated matrix elements.  A
similar problem can
arise in finite size lattice problems, where large variations in results can
persist up
to very large system sizes\cite{Cox private}.  The solution for that particular
problem is an
average over boundary conditions.  Our problem, unfortunately, has no such
cure, and we must
live with reasonable conclusions from the data we are able to gather.

To reiterate, we have been able to draw the following conclusions:

\noindent
$\bullet$ The best candidate for a pairing instability is in a state
of $T_{1g}$  ($xy(x^2+y^2)$) symmetry, with $\Gamma_{T1g}/T_{o7}$
=-0.212$\pm$0.025.

\noindent
$\bullet$ The $E_g$ channel ($x^2-y^2$, $3z^2-r^2$) also shows weak signs
of an instability, with
\newline
{\protect $\Gamma_{Eg}/T_{o7}$=-0.163$\pm$0.054}.

\noindent
$\bullet$ Quasiparticle interactions are strong and repulsize in the $A_{1g}$
(``s-wave'')
pairing channel.

\noindent
$\bullet$ A pairing instability of $T_{2g}$ ($xy$, $xz$, $yz$) symmetry also
appears highly unlikely.

\noindent
Using the classic weak-coupling equation for the superconducting transition
temperature within a given
representation, $\eta$, we find (in the $T_{1g}$ channel)
\begin{equation}  \label{eq:Tct1g}
T_c(\eta =T_{1g})=1.13T_{o7}e^{T_{o7}/\Gamma_{T1g}}\approx 0.09\hspace{0.2cm}
K,
\end{equation}
which is smaller than the measured $T_c$ of CeCu$_2$Si$_2$.  We did not expect,
however, such a
weak-coupling calculation to give a quantitatively accurate value for $T_c$,
which leads us to say a few words
about strong versus weak coupling results.

In the case of heavy Fermions, the effective Fermi
temperature is of the order of the Kondo temperature, which is also the energy
scale
of importance for the superconducting glue.  That is, on physical grounds there
is no reason
to believe that only a very thin energy shell (thin compared to the Kondo
temperature, $T_{o7}$)
about the Fermi surface is of importance for
superconductivity.  Thus, a strong-coupling calculation, including the energy
dependence of the scattering
amplitude, should be performed.  Given the
complexity of the static problem, however, including the dynamics of the slave
Bosons
in the presence of crystal fields is not feasible.
This does not necessarily imply that the pairing instabilities
found in the static problem have no meaning.  We shall discuss, generally, why
this is so.

It is conventional wisdom\cite{Varma 85}, that near the Fermi surface the
quasiparticle self-energy for heavy Fermions is strongly frequency dependent
but only weakly
dependent on the magnitude of the momentum, $|\vec k|$.  This can be understood
intuitively as
follows.  The characteristic energy scale for the quasiparticles is the Kondo
temperature,
$T_{o7}$, which is about 10K.  The degeneracy temperature for a typical metal
is $T_F\approx$ 10,000K.
The characteristic momentum, however, is set by the Fermi wavevector, $k_F$,
which for \cc is
the size of a typical metal.  Thus, broadly speaking, we expect the
quasiparticle self-energy,
$\Sigma$, to behave (near the Fermi surface) as
$$\frac{\partial\Sigma}{\partial\omega}\approx{\cal
O}\biggl(\frac{\Sigma}{T_{o7}}\biggr),$$
$$\frac{\partial\Sigma}{\partial\xi_{\vec k}}\approx{\cal
O}\biggl(\frac{\Sigma}{D}\biggr),$$
where $D$ is the bandwidth of the conduction band.  Then, since
$$\frac{\partial\Sigma}{\partial\omega}\gg\frac{\partial\Sigma}{\partial\xi_{\vec k}},$$
it seems reasonable to ignore the momentum dependence of the self-energy.  This
intuitive
result was reinforced by Millis and Lee\cite{Millis 86}, who found (in the
$SU(N)$ model) that
the momentum dependence of the imaginary part of the conduction electron
self-energy
(at order $1/N$) is very weak (going as $1/N^2$), while the frequency
dependence goes as the
inverse of the Kondo scale.

If one accepts the dominance of the frequency dependence in the quasiparticle
self-energy, then
it seems reasonable to assume that including such dependence in a strong
coupling calculation
(ala McMillan\cite{McMillan 68}) would serve only to reduce the transition
temperature, $T_c$.
We do not believe that the {\em sign} of the average scattering amplitude,
$\Gamma_{\eta}$,
would be affected by such frequency dependence.  Thus, our conclusions about
which pairing
channels, $\eta$, show a superconducting instability should not be changed as
the
result of a strong coupling calculation.

Please note that we are {\em not} saying that the reduction of the transition
temperature due to
the frequency dependence of the residual quasiparticle interactions would be
identical in
structure to the case of electron-phonon coupling\cite{McMillan 68}.  We can
only say now that
we expect $T_c$ to be reduced; we can not give an estimate of how large the
reduction would be.

The superconducting instabilities themselves appear to be based heavily on the
underlying symmetry
of the problem, which does not care if one performs a strong or weak coupling
calculation.
The fact that our estimated (weak-coupling) transition temperature for the
$T_{1g}$ pairing
instability is smaller than the measured value for \cc, is not a
surprise.   The major purpose
of this calculation has not been to give a precise
numerical recipe for calculating the $T_c$ of heavy Fermion systems.  We wished
to study
the importance of local, or ``multiplet'', physics upon quasiparticle
interactions.  Thus the
a weak coupling calculation should be a reasonable starting point.

\pagebreak
\centerline{\bf Acknowledgements}

\vspace{1.0truecm}
\noindent
$^*$ Present address:  Department of Physics, Monmouth College, Monmouth IL.

\vspace{1.0truecm}
The author wishes to thank D. L. Cox for invaluable guidance and support over
the
duration of this project.  The author also wishes to thank the following people
for meaningful discussions:  J. W. Wilkins, C. Pennington, M. Alouani, M.
Steiner,
and Q. Si.  This work was supported by a grand from the U. S. Department of
Energy,
Office of Basic Energy Sciences, Division of Materials Research.

\newpage
\vspace{2.0cm}
\begin{center}
{\Large TABLE CAPTIONS}
\end{center}
\vspace{1.0cm}

{\bf Table 1}:
  Functional forms for the angular dependence of $\mu_{\ga\gp\ap}$
  for all possible combinations of the crystal field quantum
  numbers $\Gamma, \alpha, \gp, \ap$. Note, that by time-reversal symmetry,
  $\mu_{\ga\gp\ap}$=$\mu_{\Gamma^{*}\alpha^{*}\Gamma^{\prime *}\alpha^{\prime
*}}$,
  where the * denotes the time-reversed pair.  For example,
  $\mu_{71,71}=\mu_{7,-1,7,-1}$.  Note, also, that $V_{o\Gamma}(|\vec k|)$
represents the
  dependence of the hybridization strength
  on the radial component of $\vec k$. Any combination of quantum numbers not
  present in the table or not the time-reversed pair of quantum numbers in the
table, will
  vanish upon summing over the pseudo-spin indices.

{\bf Table 2}:
    Self-consistent mean field parameter sets, labeled as (a) and (b).
    At mean field there are three coupled integral equations,
    which are solved self-consistently.  The input parameters are:  the bare
hybridization strength, $V_o$;
    the lower edge of the conduction electron band in the absence of
hybridization, -$D$;
    the total number of electrons, $n_{total}=n_{cond}
    +n_f$, per unit cell; the bare, or unshifted, energy of the $\Gamma_7$
($E_7$) and the
    $\Gamma_8$ ($E_8=E_7+\Delta_{CEF}$) multiplets; and the fixed crystal field
splitting, $\Delta_{CEF}
    =360$ K.  The self-consistent parameters which solve the equations are:
the
    hybridization renormalization coefficient, $s_o$, where the mean field
renormalized hybridization
    is $s_oV_o$;  the shifted $\Gamma_7$ ($\epsilon_7$) and $\Gamma_8$
($\epsilon_8=\epsilon_7+\Delta_{CEF}$)
    multiplet energies; the {\em quasiparticle} chemical potential, $\mu$; and
the Kondo temperature
    $T_{o7}\equiv\epsilon_7-\mu$ of the $\Gamma_7$ doublet.
    In both parameters sets (a) and (b), the total number of particles was
    fixed at $n_{total}$=1.5.  And the input parameters were chosen to give
approximate Kondo
    temperatures of 10 K, {\em i.e.} $T_{o7}\approx$10 K.  {\em All energies
are measured relative
    the chemical potential in the absence of hybridization}.  In both parameter
sets,
    $\Delta_{CEF}$=360 K, and the unshifted $\Gamma_7$ energy is $E_7$=-2.0 eV.

{\bf Table 3}:
  The parameters $x_{\Gamma}$ and $y_{\Gamma}$ are for both mean field
parameter sets described
  in Table 2.

{\bf Table 4}:
  Realizations of the cubic harmonics, $\Phi_{\eta}$,  as linear combinations
  of the spherical harmonics $Y_{lm}$.
  For each representation, $\eta$,  of cubic symmetry, the expansion was cut
off after the lowest set of
  spherical harmonics with $l>0$.

{\bf Table 5}:
  Irreducible representations of the octahedral group, $O$.  To get the group
$O_h$, we add inversions
  to the allowed symmetry operations, the result of which is that all
representations pick up a
  subscript, $g$ (for even parity) or $u$ (for odd parity).

{\bf Table 6}:
   The local (``hard-core'') quasiparticle scattering amplitude in the presence
of
   crystal electric fields, $\Gamma_{QP}$,
   averaged over a spherical Fermi surface.  Results for both mean field
   parameter sets (a) and (b) are given. See table 5.3 for a discussion of the
mean field
   parameters themselves.  The first column gives the representations, labeled
by $\eta$,
   of the group $O_h$.  The second and third columns are the averaged local
scattering
   amplitudes (divided by the $\Gamma_7$ Kondo temperature, $T_{o7}$) for the
parameter sets (a) and (b),
   respectively.  The fourth column lists (for parameter set (a)) the
contribution to the average
   from the so-called hot-spots, where the Brillouin zone axes intersect the
Fermi surface.
   In column four, in the $E_g$ channel, the attractive interaction seems to be
due to these
   hot-spots; but in the $A_{1g}$ channel the hot-spots do not dominate, since
the full Fermi
   surface average is large and positive.
   The average in the $E_g$ {\em should} be
   most sensitive to the hot-spots, since that is where the $E_g$ cubic
harmonics have their
   maximum value.  In the $T_{1g}$ and $T_{2g}$ channels there is rigorously
zero contribution
   from the hot-spots because the $T_{1g}$ and $T_{2g}$ cubic harmonics vanish
at those points.

{\bf Table 7}:
  The local (``hard-core'') quasiparticle scattering amplitude in the jellium
model,
  as studied by Zhang and Lee[1].
  When divided by the Kondo temperature and averaged over the spherical
  Fermi surface, these results are {\em universal};  there are no other
parameters involved.
  In spherical symmetry, the pairing states are labeled by their relative
angular
  momentum, $l$, with $l$=0 corresponding the s-wave; $l$=2, d-wave; and $l$=4,
g-wave.
  It is meaningful to compare Zhang and Lee's results with ours, because in the
limit of
  spherical symmetry, where all the Bosonic propagators in equation 5.37 are
replaced with
  their values in spherical symmetry, and where the normalization functions,
$A^{2}_{1}(\vec k)$
  and $A^{2}_{1}(\kp)$, are replaced by their isotropic values in spherical
symmetry, our
  expression for $\Gamma_{QP}$ (equation 5.73) gives exactly the same local
interactions
  as Zhang and Lee.  Note that all local interactions are {\em repulsive} in
spherical symmetry.

\newpage
\vspace{2.0cm}
\begin{center}
{\Large FIGURE CAPTIONS}
\end{center}
\vspace{1.0cm}

{\bf Figure 1}:
  Splitting of $J=5/2$ multiplet into a $\Gamma_7$ doublet and a $\Gamma_8$
quartet
  due to crystal fields of cubic symmetry.  Neutron scattering gives a
splitting of
  about 360 K.

{\bf Figure 2}:
  A schematic representation of the hybridization process, in which a
conduction
  electron in the state $|$$\vec k \sigma$$>$ jumps into an empty $4f$ orbital
of total
  angular momentum $J=5/2$, and z-component $m$
  described by $|$m$>$.  The matrix element is $<$m$|$$\hat V$$|$$\vec
k\sigma$$>$=
  V$_{m\sigma}$($\vec k$).  If we expand the conduction state in partial waves,
then
  only the state with total angular momentum $J=5/2$ and z-component $m$ would
  be able to hybridize.  Note that the energy of the $J$=7/2 excited multiplet
  is physically too large; it is shown thus merely to make the figure more
  readable.  The actual size of the spin orbit coupling in \cc, should be
  about an order of magnitude larger than the crystal field splitting.  Note,
  also, that the energy splittings of the $\Gamma_6$, $\Gamma_7$, and
$\Gamma_8$
  states, which come from the crystal field splitting of the $J$=7/2
  multiplet, are not accurate.  The states are shown merely to inform the
  reader how the $J$=7/2 multiplet decomposes in cubic symmetry.
  See, for example, Appendix D in reference [40].

{\bf Figure 3}:
  Schematic of the quasiparticle bandstructure, showing the shifted crystal
field multiplet
  energies ($\epsilon_7$ and $\epsilon_8$), the chemical potential ($\mu$), and
the
  lower band edge (-D).  (a) the k values range from the zone center (the
$\Gamma$ point) to
  the intersection of the $k_x$ axis with the cubic Brillouin zone boundary.
(b) From the
  zone center to the intersection of a cube diagonal with the Brillouin zone
boundary.

{\bf Figure 4}:
   A plot of the normalization function $A^{2}_{1}(\vec k)$ along the equator
of a sphere
   in k-space as a function of azimuthal angle, $\phi$. The sharp spikes occur
at the
   intersections of the equator with the coordinate axes.  At these points the
   $\Gamma_7$ hybridization matrix elements vanish exactly, and the width of
these peaks is
   set by the ratio of the two Kondo temperatures $T_{o7}/T_{o8}\approx
T_{o7}/\Delta_{CEF}$.
   The behavior for a fixed azimuthal angle ($\phi$=0) as a function of the
polar angle, $\theta$,
   is the same.

{\bf Figure 5}:
  Plot of the lowest energy band, $E_{1\vec k}$, for the two mean-field
parameter sets (labeled as
  (a) and (b)).  In set (a), the conduction electron chemical potential is
2.44\hspace{0.2cm}eV, and
  in set (b), it is 3.33\hspace{0.2cm}eV.  Note that the bands are just shifted
with respect to each
  other.
  For both parameter sets the unshifted $\Gamma_7$ multiplet energy sits at
-2.0\hspace{0.2cm}eV.

{\bf Figure 6}:
  A diagrammatic representation of the infinite-order summation that gives the
mean-field
  dressed f-electron propagator.  All the Green functions are defined
pictorially at
  the top of the figure.  The bare conduction electron propagator is
represented by a
  solid line.

{\bf Figure 7}:
  Diagrammatic representation of Dyson's equation for the conduction Green
function.  The
  double line is the hybridization dressed conduction Green function, and
$\Sigma$$_{c\sigma}$($\vec k$)
  is the conduction self-energy.  All other elements of the figure are as
defined in
  Figure 6.

{\bf Figure 8}:
  Diagrammatic representation of Dyson's equation for the mixing Green
function.

{\bf Figure 9}:
  (a). Dyson's equation
  for the diagonal component (in the crystal field indices) of the inverse of
the (matrix)
  Bosonic propagator $\hat D^{-1}_{\Gamma\Gamma}$.
  (b). The three hybridization-dressed (mean-field) Fermionic Green functions.
Note that, in general,
  the $f$ Green function can mix the multiplet indices at k-points away from
the zone center.
  This is because away from the zone center, the symmetry is lower than cubic,
allowing the
  crystal field indices to mix.
  (c). Leading contributions to the components ($ss, s\lambda,$ and
$\lambda\lambda$) of the
  self-energy matrix from closed Fermionic loops.  The ``$\times$'' symbol
represents the
  scaled hybridization matrix element, $\tilde V_{\ga\sigma}(\vec k)$, which is
assumed to be
  of order one.  The external legs in the $\Gamma_{ss\Gamma\Gamma}$ diagram
come from the
  scaled fluctuations in the $s$ fields, $\delta\tilde s_{\vec q\Gamma}$.


{\bf Figure 10}:
  Numerically evaluated function $I_{ss}(\vec q)$ as a function of the mesh
parameter, $n_{mesh}$
  for mean-field parameter set (a).
  Here we used $\vec q =0.5\hat z$.

{\bf Figure 11}:
  Numerically evaluated function $I_{s\lambda}(\vec q)$ as a function of the
mesh parameter, $n_{mesh}$
  for mean-field parameter set (a).
  Here we used $\vec q =0.5\hat z$.

{\bf Figure 12}:
  Numerically evaluated function $I_{\lambda\lambda}(\vec q)$ as a function of
the mesh parameter, $n_{mesh}$
  for mean-field parameter set (a).
  Here we used $\vec q =0.5\hat z$.

{\bf Figure 13}:
  Quasiparticle scattering amplitude for incoming particles (solid lines) of
momenta $\vec k$ and
  -$\vec k$.  The wavy lines are dressed boson propagators, and the vertices
  denote the anisotropic coupling of quasiparticles to bosons.

{\bf Figure 14}:
   Fermi surface averages in the $A_{1g}$ and $T_{2g}$ pairing channels of the
   scattering amplitude, $\langle\Gamma\rangle/T_{o7}$, as a function of the
Boson
   mesh parameter  $n_{mesh}$.  8$n^{3}_{mesh}$ equals the number of tetrahedra
   used in the Brillouin zone integrals for the Bosonic Green functions.
   These data are for mean field parameter set (a).
   (See Table 5.3.)\newline
   \noindent
   $\bullet$The horizontal line marked by $T_{2g,local}$ denotes the
   size of the local ("hard-core") contribution to the average in the $T_{2g}$
channel.  The corresponding
   local contribution for the $A_{1g}$ channel is 3.54 and
   would be just above the top of
   the graph.  We see that the nonlocal contribution in the $T_{2g}$ channel is
of the
   same size as the local contribution.  Even with the fluctuation at
$n_{mesh}$=23,
   it seems unlikely there is a pairing instability of $T_{2g}$
symmetry.\newline
   \noindent
   $\bullet$In the $A_{1g}$ channel, even with the large fluctuation for
$n_{mesh}$=39,
   there is clearly no pairing instability.

{\bf Figure 15}:
   Fermi surface averages in the $E_{1g}$ and $T_{1g}$ pairing channels of the
   scattering amplitude, $\langle\Gamma\rangle/T_{o7}$, as a function of the
Boson
   mesh parameter  $n_{mesh}$.  These data are for mean field parameter set
(a).
   The solid (dashed) horizontal line denotes the local contribution
   to the average in the $E_g$ ($T_{1g}$) channel.\newline
   \noindent
   $\bullet$Although not yet converged, the averages in the $T_{1g}$ channel
point to the
   possibility of a pairing instability.  Averaging the values of
$\langle\Gamma\rangle
   _{T1g}/T_{o7}$ for all the values of $n_{mesh}$ used here yields
   $$ \frac{\langle\Gamma\rangle_{T1g}}{T_{o7}}=-0.212\pm 0.025.$$\newline
   \noindent
   $\bullet$In the $E_g$ channel, there is a large fluctuation at
$n_{mesh}$=39.  Therefore,
   we are hesitant to say that this is evidence of a superconducting
instability.
   However, it is clear that the average interactions in this
   channel are attractive for a relatively wide range of mesh sizes: 25$\leq
n_{mesh}\leq$ 47.

{\bf Figure 16}:
   Fermi surface averages in the $A_{1g}$ and $T_{2g}$ pairing channels of the
   scattering amplitude, $\langle\Gamma\rangle/T_{o7}$, as a function of the
Boson
   mesh parameter  $n_{mesh}$. (8$n^{3}_{mesh}$ equals the number of tetrahedra
used in
   the Brillouin zone integrals for the Bosonic Green functions.)
   These data are for mean field parameter set (b).\newline
   \noindent
   $\bullet$In the $T_{2g}$ channel, the horizontal, dashed line denotes the
   local contribution to the average.  The fluctuations in the full average are
at least a
   factor of two larger than the local part and are also varying about zero.
Thus it is
   not possible to say, from the present data, if there is a $T_{2g}$
instability or not.
   Using the results from parameter set (a), however, it still seems unlikely
that there
   is an instability in this channel.\newline
   \noindent
   $\bullet$In the $A_{1g}$ channel, it is easy to see that there average
interactions are
   repulsive and strong.  There is no instability in this channel.

{\bf Figure 17}:
   Fermi surface averages in the $E_{1g}$ and $T_{1g}$ pairing channels of the
   scattering amplitude, $\langle\Gamma\rangle/T_{o7}$, as a function of the
Boson
   mesh parameter  $n_{mesh}$.  These data are for mean field parameter set
(b).\newline
   \noindent
   $\bullet$In the $E_g$ channel, the horizontal line denotes the contribution
to the average
   in the local limit.  For the full average, up to $n_{mesh}$=31, the average
value is
   fluctuating evenly about the local value.  The surprisingly large
fluctuation at
   $n_{mesh}$=31, however, makes it impossible to tell if there is an
instability in this channel.
   \newline
   \noindent
   $\bullet$In the $T_{1g}$ channel, the full average, although fluctuating,
remains negative
   for 21$\leq n_{mesh}\leq$ 41.  Averaging these values gives a result of
   $$\frac{\langle\Gamma\rangle_{T1g}}{T_{o7}}=-0.226\pm 0.049.$$

\newpage
\begin{tabular}{|c|c|c|}\hline
$\Gamma,\alpha$ & $\gp,\ap$ & $\mu_{\Gamma\alpha\gp\ap}(\vec k)/\tilde
V_{o\Gamma}(|\vec k|)
\tilde V_{o\gp}(|\vec k|)$ \\ \hline
7,1 & 7,1 &
$-\frac{2\sqrt{\pi}}{3}\bigl[Y_{00}-\frac{1}{3}Y_{40}-\frac{1}{3}\sqrt{\frac{5}{14}}(
Y_{44}+Y_{4-4})\bigr]$ \\ \hline
8,2 & 8,2 &
$-\frac{2\sqrt{\pi}}{3}\bigl[Y_{00}-\frac{8}{7}\sqrt{\frac{1}{5}}Y_{20}+\frac{1}{21}
Y_{40}+\frac{1}{3}\sqrt{\frac{5}{14}}(Y_{44}+Y_{4-4})\bigr]$ \\ \hline
8,1 & 8,1 &
$-\frac{2\sqrt{\pi}}{3}\bigl[Y_{00}+\frac{8}{7}\sqrt{\frac{1}{5}}Y_{20}+\frac{2}{7}
Y_{40}\bigr]$ \\ \hline
8,2 & 8,1 &
$\frac{10}{21}\sqrt{\frac{2\pi}{5}}\bigl[Y_{22}+\frac{3}{5}Y_{2-2}-\frac{\sqrt{3}}{2}Y_{42}
+\frac{\sqrt{3}}{6}Y_{4-2}\bigr]$ \\ \hline
8,1 & 8,2 &
$\frac{10}{21}\sqrt{\frac{2\pi}{5}}\bigl[Y_{2-2}+\frac{3}{5}Y_{22}-\frac{\sqrt{3}}{2}Y_{4-2}
+\frac{\sqrt{3}}{6}Y_{42}\bigr]$ \\ \hline
8,2 & 8,-1 &
$-\frac{4}{21}\sqrt{\frac{2\pi}{5}}\bigl[Y_{2-1}+\frac{5}{6}\sqrt{\frac{3}{2}}Y_{4-1}
-\frac{5}{6}\sqrt{\frac{21}{2}}Y_{43}\bigr]$ \\ \hline
8,-1 & 8,2 &
$\frac{4}{21}\sqrt{\frac{2\pi}{5}}\bigl[Y_{21}+\frac{5}{6}\sqrt{\frac{3}{2}}Y_{4-1}
-\frac{5}{6}\sqrt{\frac{21}{2}}Y_{4-3}\bigr]$ \\ \hline
7,1 & 8,2 &
$\frac{4}{21}\sqrt{6\pi}\bigl[Y_{2-1}-\sqrt{\frac{1}{6}}Y_{4-1}\bigr]$ \\
\hline
8,2 & 7,1 &
$-\frac{4}{21}\sqrt{6\pi}\bigl[Y_{21}-\sqrt{\frac{1}{6}}Y_{41}\bigr]$ \\ \hline
7,1 & 8,1 &
$-\frac{4}{21}\sqrt{2\pi}\bigl[Y_{21}+\frac{1}{6}\sqrt{\frac{21}{2}}Y_{4-3}
+\frac{5}{6}\sqrt{\frac{3}{2}}Y_{41}\bigr]$ \\ \hline
8,1 & 7,1 &
$-\frac{4}{21}\sqrt{2\pi}\bigl[Y_{2-1}+\frac{1}{6}\sqrt{21}{2}Y_{43}+\frac{5}{6}
\sqrt{\frac{3}{2}}Y_{4-1}\bigr]$ \\ \hline
7,1 & 8,-1 &
$\frac{2}{21}\sqrt{18\pi}\bigl[Y_{22}-\frac{1}{3}Y_{2-2}+\frac{1}{2\sqrt{3}}Y_{4-2}
+\frac{5}{6\sqrt{3}}Y_{42}\bigr]$ \\ \hline
8,-1 & 7,1 &
$\frac{2}{21}\sqrt{18\pi}\bigl[Y_{2-2}-\frac{1}{3}Y_{22}+\frac{1}{2\sqrt{3}}Y_{42}
+\frac{5}{6\sqrt{3}}Y_{4-2}\bigr]$ \\ \hline
7,1 & 8,-2 &
$-\frac{4}{21}\sqrt{\pi}\bigl[Y_{20}-\frac{\sqrt{5}}{3}Y_{40}+\frac{5}{6}\sqrt{
\frac{7}{2}}Y_{4-4}-\frac{1}{6}\sqrt{\frac{7}{2}}Y_{44}\bigr]$ \\ \hline
8,-2 & 7,1 &
$-\frac{4}{21}\sqrt{\pi}\bigl[Y_{20}-\frac{\sqrt{5}}{3}Y_{40}+\frac{5}{6}\sqrt{\frac
{7}{2}}Y_{44}-\frac{1}{6}\sqrt{\frac{7}{2}}Y_{4-4}\bigr]$ \\ \hline
\end{tabular}

\newpage
\begin{tabular}{|c|c|c|c|c|c|c|} \hline\hline
  parameter & $V_o$ & $s_o$ & $\epsilon_7$ & $\mu$ & $T_{o7}$ & $D$\\
   set & (eV) & & (eV) & (eV) & (K) & (eV) \\ \hline
 a & 0.8595 & 0.093472 & 0.0217648 & 0.0208399 & 9.25 & 2.4405 \\ \hline
b & 0.650 & 0.142409 & -0.8437021 & -0.844999 & 13.0 & 3.33856 \\ \hline\hline
\end{tabular}

\newpage
\begin{tabular}{|c|c|c|c|c|}\hline\hline
 parameter set & $x_7$ & $x_8$ & $y_7$ & $y_8$ \\ \hline
 a & 1.820 & 0.0717 & 0.975 & 0.04029 \\ \hline
 b & 1.729 & 0.0949 & 0.925 & 0.0498 \\ \hline\hline
\end{tabular}

\newpage
\begin{tabular}{|c|c|} \hline
$\eta$ & $\Phi_{\eta}$ \\ \hline
$A_{1g}$ &
$\frac{1}{\sqrt{2}}\biggl(Y_{00}+0.76376261Y_{40}+0.4564355(Y_{44}+Y_{4-4})\biggr)$ \\ \hline
$A_{2g}$ & $0.58630197(Y_{62}+Y_{6-2})
-0.3952847(Y_{66}+Y_{6-6})$ \\ \hline
$E_{g}$ & $Y_{20}$ \\
$E_{g}$ & $\frac{1}{\sqrt{2}}(Y_{22}+Y_{2-2})$ \\  \hline
$T_{1g}$ &
$\bigl(\frac{-0.93541435i}{\sqrt{2}}(Y_{41}-Y_{4-1})-\frac{0.353553391i}{\sqrt{2}}(Y_{43}-Y_{4-3})\bigr)$ \\
$T_{1g}$ &
$\bigl(\frac{0.93541435}{\sqrt{2}}(Y_{41}+Y_{4-1})-\frac{-0.353553391}{\sqrt{2}}(Y_{43}+Y_{4-3})\biggr)$ \\
$T_{1g}$ & $\frac{i}{\sqrt{2}}(Y_{44}-Y_{4-4})$ \\ \hline
$T_{2g}$ & $\frac{i}{\sqrt{2}}(Y_{21}-Y_{2-1})$ \\
$T_{2g}$ & $\frac{1}{\sqrt{2}}(Y_{21}+Y_{2-1})$ \\
$T_{2g}$ & $\frac{i}{\sqrt{2}}(Y_{22}-Y_{2-2})$ \\ \hline
\end{tabular}

\newpage
\begin{tabular}{ccc}
\underline{\rm Representation} & \underline{\rm Dimensionality} &
\underline{\rm Transforms like} \\
$A_1$ & 1 & $x^2+y^2+z^2$ \\
$A_2$ & 1 & $(x^2-y^2)(z^2-x^2)(y^2-z^2)$  \\
$E$ & 2 & $x^2-y^2, 3z^2-r^2$ \\
$T_1$ & 3 & $x$, $y$, $z$\\
$T_2$ & 3 & $xy$, $yz$, $zx$
\end{tabular}

\newpage
\begin{tabular}{|c|c|c|c|} \hline
(a) Cubic Symmetry & $\langle\Gamma_{local}\rangle_{\eta}/T_{o7}$ &
$\langle\Gamma_{local}\rangle_{\eta}/T_{o7}$ &
$\langle\Gamma_{local,hot}\rangle/T_{o7}$ \\
$\eta$ & set (a) & set (b) & set (a)\\ \hline
$A_{1g}($s$-wave)$ & 3.54 & 3.53 & -0.215\\
$E_{g}$($d_{x^2-y^2},d_{3z^2-r^2}$)& -0.0275 & -0.0453 & -0.0696\\
$T_{1g}$ & 0.0478 & 0.0695 & 0.0\\
$T_{2g}$($d_{xy},d_{yz},d_{xz}$) & 0.0487 & 0.0709 & 0.0\\ \hline
\end{tabular}

\newpage
\begin{tabular}{|c|c|} \hline
 Spherical Symmetry & $\langle\Gamma_{local}\rangle_{\eta}/T_o$ \\ \hline
$\eta$=0 (s-wave) & 1/3 \\
$\eta$=2 (d-wave) & 8/21 \\
$\eta$=4 (g-wave) & 2/7 \\ \hline
\end{tabular}


\begin{thebibliography}{99}

\bibitem{Steglich 79}
F. Steglich, J. Aarts, C. D. Bredl, W. Lieke, D. Meschede, W. Franz, and
H. Sch\"{a}fer, Phys. Rev. Lett. {\bf 43}, 1892 (1979).
\bibitem{Grewe 90}
N. Grewe and F. Steglich, in {\it Handbook on the Physics and Chemistry of
Rare-Earths}, vol 14, edited by K. A. Gschneider Jr. and L. Eyring,
(North-Holland, Amsterdam, 1990).
\bibitem{Stewart 84}
G. R. Stewart, Rev. Mod. Phys. {\bf 56}, 755 (1984).
\bibitem{Lee 86}
P. A. Lee, T. M. Rice, J. W. Serene, and J. W. Wilkins, Comments on Condens.
Matt. Phys. {\bf 12}, 99 (1986).
\bibitem{Weslau 92}
B. Welslau and N. Grewe, Ann. Physik {\bf 1}, 214 (1992).
\bibitem{Fulde 88}
P. Fulde, J. Keller, and G. Zwicknagl, Solid State Physics {\bf 41}, 1 (1988).
\bibitem{Rainer 88}
D. Rainer, Physica Scripta {\bf T23}, 106 (1988).
\bibitem{Barnes 76}
S. E. Barnes, J. Phys. F {\bf 6}, 1375 (1976).
\bibitem{Read 83a}
N. Read and D. M. Newns, J. Phys. C {\bf 16}, L1055 (1983).
\bibitem{Coleman 84}
P. Coleman, Phys. Rev. B {\bf 29}, 3035 (1984).
\bibitem{Millis 86}
A. J. Millis and P. A. Lee, Phys. Rev. B {\bf 35}, 3394 (1986).
\bibitem{Auerbach 86}
A. Auerbach and K. Levin, Phys. Rev. Lett. {\bf 57}, 877 (1986).
\bibitem{Tesanovic 86}
Z. Te\u{s}anovi\'{c} and O. T. Valls, Phys. Rev. B {\bf 34}, 1918 (1986).
\bibitem{Coleman 87}
P. Coleman, Phys. Rev. B {\bf 35}, 5072 (1987).
\bibitem{Lavagna 87}
M. Lavagna, A. J. Millis, and P. A. Lee, Phys. Rev. Lett. {\bf 58}, 266 (1987).
\bibitem{Houghton 87}
A. Houghton, N. Read, and H. Won, Phys. Rev. B {\bf 35}, 5123 (1987).
\bibitem{Zhang 87}
F. C. Zhang and T. K. Lee, Phys. Rev. B {\bf 35}, 3651 (1987).
\bibitem{Coqblin 69}
B. Coqblin and J. R. Schrieffer, Phys. Rev. {\bf 185}, 847 (1969).
\bibitem{Zou 86}
Z. Zou and P. W. Anderson, Phys. Rev. Lett. {\bf 57}, 2073 (1986).
\bibitem{Read 83}
N. Read and D. M. Newns, J. Phys. C {\bf 16}, 3273 (1983).
\bibitem{Horn 81}
S. Horn, E. Holland-Moritz, M. Loewenhaupt, F. Steglich, H. Scheuer, A. Benoit,
and
J. Flouquet, Phys. Rev. B {\bf 23}, 3171 (1981).
\bibitem{Bredl 85}
C. D. Bredl, W. Lieke, R. Schefzyk, M. Lang, U. Rauchschwalbe, F. Steglich, S.
Riegel,
R. Felten, G. Weber, J. Klaasse, J. Aarts, and F. R. de Boer, J. Magn. Magn.
Mat. {\bf 47 \& 48},
30 (1985).
\bibitem{Steglich 85}
F. Steglich, U. Rauchschwalbe, U. Gottwich, H. M. Mayer, G. Sparn, N. Grewe, U.
Poppe, and
J. J. M. Franse, J. Appl. Phys. {\bf 57}, 3054 (1985).
\bibitem{Rauchschwalbe 87}
U. Rauchschwalbe, Physica {\bf 147B}, 1 (1987).
\bibitem{Lea 62}
K. R. Lea, M. J. M. Leask, and W. P. Wolff, J. Phys. Chem. Solids {\bf 23},
1381 (1962).
\bibitem{Kang 90}
J. S. Kang, J. W. Allen, O. Gunnarsson, N. E. Christensen, O. K. Andersen, Y.
Lassailly, M. B. Maple,
and M. S. Torikachvili, Phys. Rev. B {\bf 41}, 6610 (1990).
\bibitem{Parks 84}
R. D. Parks, M. L. den Boer, S. Raaen, J. L. Smith, and G. P. Williams, Phys.
Rev. B {\bf 30},
1580 (1984).
\bibitem{Cox Quad}
D. L. Cox, OSU preprint (1990).
\bibitem{coxprivate}
D. L. Cox, private communication.
\bibitem{Martin 82}
R. M. Martin, Phys. Rev. Lett. {\bf 48}, 362 (1982).
\bibitem{Fulde}
see. {\em e.g.}, P. Fulde, {\em Electronic Correlations in Molecules and
Solids},
(Springer-Verlag, Berlin, 1991), p. 44.
\bibitem{Anderson 84}
P. W. Anderson, Phys. Rev. B {\bf 30}, 1549, 4000 (1984).
\bibitem{Zhang 86}
F. C. Zhang and T. K. Lee, Phys. Rev. B {\bf 34}, 8114 (1986).
\bibitem{Rice 85}
T. M. Rice and K. Ueda, Phys. Rev. Lett. {\bf 55}, 995 (1985).
\bibitem{Cox 87}
D. L. Cox, Phys. Rev. Lett. {\bf 59}, 1240 (1987).
\bibitem{Zhang 87a}
F. C. Zhang and T. K. Lee, Phys. Rev. Lett. {\bf 58}, 2728 (1987).
\bibitem{Varma 87}
G. Aeppli and C. M. Varma, Phys. Rev. Lett. {\bf 58}, 2729 (1987).
808 (1987).
\bibitem{Batlogg 84}
B. Batlogg, J. P. Remeika, A. S. Cooper, and Z. Fisk, J. Appl. Phys. {\bf 55},
2001 (1984).
\bibitem{Jepsen 71}
O. Jepsen and O. K. Andersen, Solid State Commun. {\bf 9}, 1763 (1971).
\bibitem{Lehmann 70}
G. Lehmann, P. Rennert, M. Taut, and H. Wonn, Phys. Status Solidi {\bf 37}, K27
(1970).
\bibitem{Rath 75}
J. Rath and A. J. Freeman, Phys. Rev. B {\bf 11}, 2109 (1975).
\bibitem{Lingard 75}
Per-Anker Ling\aa rd, Solid State Commun. {\bf 16}, 481 (1975).
\bibitem{Jacobs 71}
see, {\em e.g.}, R. L. Jacobs and D. Lipton, in {\em Computational Methods in
Band Theory},
edited by P. M. Marcus, J. F. Janak, and A. R. Williams, (Plenum Press, New
York, 1971).
\bibitem{Trees thesis}
B. R. Trees, Ph. D. Thesis (The Ohio State University, 1993) unpublished.
\bibitem{Cox 85}
D. L. Cox, Ph. D. Thesis (Cornell University, 1985) unpublished.
\bibitem{Gilat 72}
G. Gilat, J. Comput. Phys. {\bf 10}, 432 (1972).
\bibitem{Gilat}
G. Gilat, in {\em Methods in Computational Physics}, vol. 15, edited by G.
Gilat, B. J. Alder,
S. Fernbach, and M. Rotenberg (Academic, New York, 1976).
\bibitem{Matthew}
Matthew Steiner, private communication.
\bibitem{Trees 94}
B. R. Trees and D. L. Cox, to be published in Phys. Rev. B, Rapid
Communications.
\bibitem{Asayama 88}
K. Asayama, Y. Kitaoka, and Y. Kohori, J. Magn. Magn. Mat. {\bf 76 \& 77}, 449
(1988).
\bibitem{Ueda 85}
K. Ueda and T. M. Rice, Phys. Rev. B {\bf 31}, 7114 (1985).
\bibitem{Altmann 65}
S. L. Altmann and A. P. Cracknell, Rev. Mod. Phys. {\bf 37}, 19 (1965).
\bibitem{Cox private}
K. -H. Luk and D. L. Cox, Phys. Rev B {\bf 41}, 4456 (1990).
\bibitem{Varma 85}
C. M. Varma, Phys. Rev. Lett. {\bf 55}, 2723 (1985).
\bibitem{McMillan 68}
W. L. McMillan, Phys. Rev. {\bf 167}, 331 (1968).
\end{thebibliography}
\end{document}